\documentclass{emulateapj}
\usepackage{lscape}
\newcommand{\lapprox }{{\lower0.8ex\hbox{$\buildrel <\over\sim$}}}
\newcommand{\gapprox }{{\lower0.8ex\hbox{$\buildrel >\over\sim$}}}

\slugcomment{DRAFT \today}
\shorttitle{ChESS}
\shortauthors{Covey \& Ag{\" u}eros et al.}

\begin{document}

\title{The ChaMP Extended Stellar Survey (ChESS): Photometric and Spectroscopic Properties of Serendipitously Detected Stellar X-ray Sources\altaffilmark{1}}

\author{K.~R.~Covey\altaffilmark{2,3,4}, M.~A.~Ag{\"u}eros\altaffilmark{2,5}, P.~J.~Green\altaffilmark{3}, D.~Haggard\altaffilmark{6}, W.~A.~Barkhouse\altaffilmark{7}, J.~Drake\altaffilmark{3}, N.~Evans\altaffilmark{3}, V.~Kashyap\altaffilmark{3}, D.-W.~Kim\altaffilmark{3}, A.~Mossman\altaffilmark{3}, D.~O.~Pease\altaffilmark{8}, J.~D.~Silverman\altaffilmark{9}}

\altaffiltext{1}{Observations reported here were obtained at the MMT Observatory, a joint facility of the Smithsonian Institution and the University of Arizona.}
\altaffiltext{2}{The first two authors contributed equally to this study.} 
\altaffiltext{3}{Harvard-Smithsonian Center for Astrophysics, 60 Garden Street, Cambridge, MA 02138} 
\altaffiltext{4}{Spitzer Fellow}
\altaffiltext{5}{NSF Astronomy and Astrophysics Postdoctoral Fellow; Columbia University, Department of Astronomy, 550 West 120th Street, New York, NY 10027} 
\altaffiltext{6}{NASA Harriett G.\ Jenkins Predoctoral Fellow, University of Washington, Department of Astronomy, Box 351580, Seattle, WA 98195} 
\altaffiltext{7}{Physics Department, University of North Dakota, Grand Forks, ND 58202} 
\altaffiltext{8}{Space Sciences Lab, 7 Gauss Way, Berkeley, CA 94720-7450}
\altaffiltext{9}{Max-Planck-Institut f\"ur extraterrestrische Physik, D-84571 Garching, Germany}

\begin{abstract}
We present $348$ X-ray emitting stars identified from
correlating the Extended {\it Chandra} Multiwavelength Project
(ChaMP), a wide-area serendipitous survey based on archival X-ray
images, with the Sloan Digital Sky Survey (SDSS). We use morphological 
star/galaxy separation, matching to an SDSS quasar catalog, an
optical color-magnitude cut, and X-ray data quality tests to create our catalog,
the ChaMP Extended Stellar Survey (ChESS), from a sample of $2121$ 
matched ChaMP/SDSS sources.  Our cuts retain $92\%$ of the spectroscopically 
confirmed stars in the original sample while excluding $99.6\%$ of the $684$
spectroscopically confirmed extragalactic sources. Fewer than $3\%$ of
the sources in our final catalog are previously identified stellar
X-ray emitters. For $42$ catalog members, spectroscopic
classifications are available in the literature. We present new
spectral classifications and H$\alpha$ measurements for an additional
$79$ stars. The catalog is dominated by main sequence stars; we estimate the 
fraction of giants in ChESS is $\sim 10\%$. We  
identify seven giant stars (including a possible Cepheid and an RR Lyrae star)
as ChAMP sources, as well as three cataclysmic variables.  
We derive distances from $\sim 10-2000$ pc for the 
stars in our catalog using photometric parallax relations appropriate for dwarfs 
on the main sequence and calculate their X-ray and bolometric luminosities.
These stars lie in a unique space in the L$_{\rm X}$--distance plane,
filling the gap between the nearby stars identified as counterparts to
sources in the {\it ROSAT} All-Sky Survey and the more distant stars
detected in deep {\it Chandra} and {\it XMM-Newton} surveys. 
For $36$ newly identified X-ray emitting M stars 
we calculate L$_{\rm H\alpha}$/L$_{\rm bol}$. L$_{\rm H\alpha}$/L$_{\rm bol}$ 
and L$_{\rm X}$/L$_{\rm bol}$ are linearly related below 
L$_{\rm X}$/L$_{\rm bol} \sim 3 \times 10^{-4}$, while 
L$_{\rm H\alpha}$/L$_{\rm bol}$ appears to turn over at larger
L$_{\rm X}$/L$_{\rm bol}$ values. 
Stars with reliable SDSS photometry have an $\sim0.1$ mag blue excess 
in $u-g$, likely due to increased chromospheric continuum emission. 
Photometric metallicity estimates suggest that the sample is evenly split 
between the young and old disk populations of the Galaxy; the lowest 
activity sources belong to the old disk population, a clear signature of the
decay of magnetic activity with age. Future papers will present
analyses of source variability and comparisons of this catalog to
models of stellar activity in the Galactic disk.
\end{abstract}


\section{Introduction}

While X-ray source counterparts are now known to range from distant
quasars to nearby active M dwarfs \citep[e.g.,][]{stocke83, stocke91,
schmitt95, zickgraf03, green04, anderson06}, X-ray data alone are
frequently insufficient to determine unambiguously whether a given
source is Galactic or extragalactic, or to make finer distinctions
about its nature. Campaigns to find optical counterparts to X-ray
sources have therefore been natural companions to the creation of
X-ray source lists since the days of the {\it Einstein Observatory}.

The Medium Sensitivity Survey \citep[MSS;][]{gioia84} and Extended
Medium-Sensitivity Survey \citep[][]{gioia90} both required
painstaking programs to identify counterparts to sources
serendipitously detected in {\it Einstein} observations. To find 
counterparts to $63$ of the $112$ MSS sources, \citet{stocke83}
obtained spectra for all of the optical objects inside or just outside
the X-ray $90\%$ confidence positional error circles--areas of radius
$\sim 30$\arcsec\ to $70$\arcsec. Once they found a plausible
counterpart by comparing its $f_X / f_V$ to that of similar objects
detected in pointed {\it Einstein} observations, \citet{stocke83}
continued to collect spectra until they reached objects at least four
times fainter than the proposed counterpart or the $\sim20.5$ mag
limit of the Palomar Observatory Sky Survey (POSS). They found that
$\sim 25\%$ of MSS sources were coronally emitting stars, primarily
late-type dwarfs; they also found one cataclysmic variable (CV).

Similar efforts have been undertaken to identify some of the
$\sim125,000$ sources included in the {\it ROSAT} All-Sky Survey
(RASS) Bright and Faint Source Catalogs \citep[BSC and
FSC;][]{voges99,fsc}. Only a relatively small fraction of RASS sources
can be identified from correlations to existing
databases. \citet{bade98} found that $35\%$ of the $80,000$ RASS
sources they considered had counterparts in SIMBAD and the NASA/IPAC
Extragalactic Database. To identify other BSC sources, \citet{bade98}
used objective prism spectra obtained as part of the Hamburg Quasar
Survey \citep[HQS;][]{hagen95} and found candidate counterparts for
$81.2\%$ of the $3847$ sources within the HQS footprint\footnote{The
unidentified sources are likely to be faint active galactic nuclei and
clusters \citep{bade98}.}. $155$ ($4\%$) are M stars, $136$ ($3.5\%$)
K stars, and $4$ ($0.1\%$) F or G stars. Another $956$ ($24.9\%$) are
saturated stars ($B \leq 14$ mag) for which no spectral class is
available. There are also $31$ white dwarfs (WDs; $0.8\%$) and $16$
CVs ($0.4\%$). There are uncertainties associated with these
identifications, e.g., because of the resolution of the spectra (R
$\approx 100$ at H$\gamma$). But the RASS/HQS program suggests that
$\sim33\%$ of the X-ray sources detected by {\it ROSAT} are Galactic
stars, a result confirmed by later efforts
\citep[e.g.,][]{zickgraf03}.

The {\it Chandra X-ray Observatory} and the {\it XMM-Newton X-ray
Observatory} are both equipped with more sensitive X-ray detectors
than {\it ROSAT} (albeit in different energy bands), but were designed
primarily to conduct pointed observations. However, growing data
archives have enabled a number of fairly deep, relatively small-area
surveys, with X-ray source lists assembled and optical counterparts
identified in much the same way as for the {\it Einstein} surveys. In
addition, a few deep pencil-beam surveys have been completed with
{\it Chandra} and {\it XMM-Newton}. \citet{Brandt2005} compare the
flux limits and solid angles for a number of these surveys; see their
Figure 1.

The selection of optical counterparts for follow-up spectroscopy is
generally simpler in these more recent surveys: the X-ray positional
uncertainties are very small (typically less than
$1$\arcsec\ for {\it Chandra}). However, the focus of these surveys is
often to characterize faint extragalactic X-ray emitters, and the
stellar samples they provide are quite small.

For example, the {\it XMM} Bright Serendipitous Survey
\citep[BSS;][]{dellac04} includes just under $400$ sources. The BSS
reaches a flux limit of $\sim7 \times 10^{-14}$ erg cm$^{-2}$ s$^{-1}$
in the $0.5-4.5$ keV energy band for an area of
$28.10$~deg$^2$. $90\%$ of the optical counterparts have magnitudes
brighter than the POSS II limit of $R\sim21$ mag \citep{dellac04}, and
close to $90\%$ of these counterparts now have spectra
\citep{lopez07}. Of these, \citet{lopez07} identified $58$ as stars,
which therefore constitute $\sim15\%$ of the X-ray counterparts--a
smaller fraction than in the {\it Einstein} or {\it ROSAT} samples,
but one which is consistent with the positions on the sky of the BSS
fields, which are $>20$~deg from the Galactic Plane. These authors
compare the colors of their $58$ stars to those predicted by the X-ray
Galactic model XCOUNT \citep{favata92}. They find that model and data
agree fairly well for the M stars in the sample but disagree
rather dramatically for F, G, and K stars. They infer that the
discrepancy is due to a stellar population currently absent from their
model, possibly known X-ray emitting binaries such as RS CVn or BY
Dra systems.

\citet{feigelson04} collected a smaller stellar sample from the {\it
Chandra} Deep Field-North (CDF-N) survey. The CDF-N has an area of
$\sim448$~arcmin$^2$; individual exposures were as long as
$\sim2\times10^6$ s, resulting in a flux limit of $3 \times 10^{-17}$
erg cm$^{-2}$ s$^{-1}$ in the $0.5-2.0$ keV band
\citep{alexander2003}. Of the $\sim500$ sources in the CDF-N, only
$\sim3\%$ are stars, and \citet{feigelson04} use $11$ of these to
construct a statistically complete sample and study the evolution of
X-ray properties. These stars belong primarily to an old-disk
population (ages between $3$ and $11$ Gyr), and their X-ray properties
are consistent with a faster-than-expected decline in magnetic
activity \citep[log L$_{\rm X} \propto t^{-2}$ rather than $t^{-1}$,
where $t$ is age;][]{feigelson04}.

Studies such as these would clearly benefit from a larger sample of
X-ray emitting stars to analyze. The {\it XMM} Slew Survey
\citep{freyberg06}, constructed from $\leq 15$~s exposures as the
satellite slewed, is one such survey. The recently released XMMSL1
catalog covers $\sim5800$ deg$^2$ to a relatively shallow flux limit
of $6\times10^{-13}$ erg s$^{-1}$ cm$^{-2}$ and includes $2692$
sources in its ``clean'' version \citep{saxton2008}. A search of the
currently available XMMSL1 database finds that $410$ {\it XMM} sources
have a star cataloged in SIMBAD within $6$\arcsec, and it is clear
that this program will eventually yield a large number of stellar
X-ray sources. However, this stellar sample is still largely
undefined. For example, re-matching the $410$ sources to SIMBAD
reveals that $35\%$ have previously been identified as RASS
sources. More work is necessary before we know exactly how many {\it
new} stellar X-ray sources will come from this survey, or the similarly
serendipitous 2XMM survey \citep{Watson2006}.

We have collected the largest sample of stellar X-ray emitters in the 
field of the Galaxy identified and characterized to date from {\it Chandra} 
or {\it XMM} data. The X-ray data are from the Extended {\it Chandra}
Multiwavelength Project (ChaMP), considerably easing the challenge of
identifying the X-ray sources. {\em Chandra} provides
sub-arcsecond astrometry over most of its field of view
\citep{aldcroft2000}, greatly facilitating unambiguous matching to
optical counterparts, as does the lack of crowding at the high
Galactic latitudes of the survey ($|b|>20$ deg). In addition, the Extended
ChaMP survey is designed to have significant overlap with the Sloan
Digital Sky Survey (SDSS), which affords well-calibrated multi-color
imaging and spectroscopy crucial both for elimination of extragalactic
objects and for classification of stars.

We describe the ChaMP and SDSS in \S\ref{surveys}, and the process by
which we identify candidate stellar counterparts in \S\ref{id}. In
\S\ref{pure} we discuss the various tests we use to confirm that these
candidates are in fact stellar X-ray emitters. In \S\ref{science} we
analyze the properties of our resulting sample of $348$ X-ray emitting
stars; we conclude in \S\ref{concl}. Future work will analyze the
X-ray variability of these stars and compare the properties of this
catalog to stellar population models of the Galaxy incorporating
evolution of time-dependent coronal X-ray emission.

\section{The Surveys}\label{surveys}

\subsection{The Extended Chandra Multiwavelength Project}\label{champx}

The {\it Chandra} Multiwavelength Project (ChaMP) is a wide-area
serendipitous survey based on archival X-ray images of the $|b|>20$
deg sky observed with the Advanced CCD Imaging Spectrometer (ACIS) on
board {\it Chandra} \citep[described in][]{weisskopf02}. The full
130-field Cycle\,1--2 X-ray catalogs are public
\citep{DKim04a,MKim07a}, and the most comprehensive X-ray number
counts (log $N$-log $S$) to date have been produced, thanks to $6600$ 
sources and massive X-ray source-retrieval simulations
\citep{DKim04b,MKim07b}. The simulations added one thousand 
artificial X-ray point sources across a wide range of fluxes to each
actual {\it Chandra} ACIS image. The resulting images were subjected
to the identical source detection and characterization as used for the
actual survey, and a comparison of input and output properties allowed
a full calculation of the ChaMP's X-ray sky coverage and completeness
as a function of e.g., source flux and off-axis angle \citep{MKim07b}.

\citet{green04} used deep imaging ($r\sim25$ mag) with the NOAO $4$-m
telescopes at KPNO and CTIO and follow-up spectroscopy with telescopes
ranging from $1.5$ to $10$ m in diameter to obtain X-ray source
identifications over $14$ deg$^2$ of the Cycle 1--2 survey. $66$ ChaMP
fields were imaged in the $g,\,r,$ and $i$ bands; these data and
photometric catalogs are available on the ChaMP webpage\footnote{\tt
http://hea-www.harvard.edu/CHAMP/} (see also Barkhouse et al.\ 2008,
in preparation). Optical spectra to $r\sim 22$ were obtained for as
many objects as feasible in 27 prime fields, using primarily the WIYN
$3.5$~m on Kitt Peak, the MMT with the Blue Channel spectrograph on Mt
Hopkins, Arizona, and the Magellan/Baade 6$.5$-m telescope with both
the LRIS and IMACS spectrographs. A significant number of
spectroscopic identifications were also obtained for $r\sim 18$
objects using the Fred Lawrence Whipple Observatory 1.5-m telescope
with the FAST spectrograph. \citet{green04} classified $125$ X-ray
counterparts with optical spectroscopy. Of these, $90\%$ are
extragalactic in nature, as expected ($63$ are broad-line
AGN). Silverman et al.\ (2008, in press) describe the
spectroscopic effort in more detail in their paper on the AGN X-ray
luminosity function, and a full ChaMP spectroscopic catalog is in
preparation.

\begin{figure}
\plotone{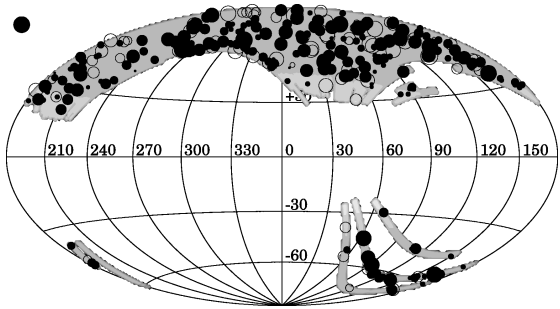}
\caption {\normalsize{The Extended ChaMP footprint in Galactic
coordinates. Open circles indicate fields observed with the ACIS-I
detector, while filled circles indicate fields observed with the
ACIS-S detector. The symbol size is proportional to the log of the
exposure time; the symbol in the upper left corner corresponds to a
$100$~ksec exposure. The SDSS footprint is the shaded
region.}}\label{fields}
\end{figure}

Given the need for even wider survey area to accumulate significant
samples of rare objects, and the time-consuming nature of deep
imaging and spectroscopy, the ChaMP area has been extended to cover
archival images from Cycles 1--6, but only to include {\it Chandra}
images within the SDSS footprint (see \S\ref{sdss}). The Extended
ChaMP now includes $392$ ACIS fields covering a total area of roughly
$33$~deg$^2$ (see Figure~\ref{fields}) and catalogs $\sim17,000$ 
X-ray sources\footnote{Some of the weakest sources may be associated with, or contaminated by, cosmic-ray afterglows. Afterglows rarely affect brighter sources, or those with bright optical counterparts as in the current sample. See also \S\ref{x_cuts}.}. The median exposure time is $21$ ksec, but individual
exposures range from $1$ to $119$ ksec. Due to the low {\it Chandra}
background rates, the formal statistical errors in net counts for each
band are consistent within 2\% of Poisson.  Here we adopt the more
conservative \citet{gehrels86} prescription: $\sigma_{cts} = 1 + (N +
0.75)^{0.5}$.

SDSS photometry within about $20$\arcmin\ of the aimpoint for
each cataloged {\it Chandra} observation were obtained to cover 
the combined ACIS-I and ACIS-S fields of view\footnote{For some observations, 
this was extended to a radius of $28$\arcmin\ to achieve full coverage of the
{\it Chandra} footprint.}. Because the {\it Chandra} point spread
function (PSF) increases with off-axis angle, comparatively few X-ray
sources are detected beyond this radius and source centroids also tend
to be highly uncertain. We note that some SDSS imaging strips do not
completely cover the {\it Chandra} field of view. 
Detailed X-ray sky coverage vs. sensitivity maps represent a major
ongoing effort of the ChaMP, described in Green et al. 2008 (in 
preparation), which will facilitate accurate volume-limit estimates
and allow for e.g., luminosity function calculations and stellar
population modeling.

While most ChaMP research to date has emphasized extragalactic objects
\citep[e.g.,][and Green et al.\ 2008, in
preparation]{Silverman05,Barkhouse06,DKim06}, the ChaMP lends itself
well to stellar research. Compared to Galactic Plane studies,
counterpart identification is very secure at the ChaMP survey's high
Galactic latitudes, crowded-field photometry is not an issue, and
reddening is quite moderate. In addition, a more balanced ratio of
thin/thick disk populations is sampled. However, the expected fraction of
stellar X-ray sources detected in the ChaMP fields is relatively low:
ChaMP fields, like those in the BSS, are away from the Plane
and stars are on average weak X-ray emitters.

\subsection{The Sloan Digital Sky Survey}\label{sdss}

The Sloan Digital Sky Survey \citep{fukugita, gunn, hogg01, smith02, gunn06} is the deepest large-scale optical survey to date, and provides uniform photometric \citep[to a depth of $r\sim22.5$ and an accuracy of $\sim 0.02$ mag;][]{zeljko04} and spectroscopic (R $\sim 1800$) datasets with which to identify ChaMP sources. The latest data release \citep[DR6;][]{DR6paper} includes imaging for $\sim9600$~deg$^2$ and photometry for close to $3\times10^8$ unique objects. The SDSS spectroscopic footprint is smaller ($\sim7400$~deg$^2$); spectra over the $3800-9200$ \AA\ range are available for $>10^6$ objects. The main spectroscopic samples are for galaxies with Petrosian $r<17.77$ ($>790,000$ objects) and quasars with PSF $i<19.1$ ($>100,000$ objects). The DR6 database also includes spectra for close to $300,000$ stars, of which nearly $70,000$ are of spectral type M or later.

SDSS photometry and spectroscopy has been used to systematically
identify RASS sources \citep[e.g.,][Ag\"ueros et al.\ 2008,
submitted]{popesso04, anderson06, Parejko2008}. While the ChaMP is a
very different survey from the RASS, the SDSS data are equally useful
in identifying ChaMP sources, and particularly stellar
sources. Typical classes of X-ray emitters, including coronally
emitting stars, normal galaxies, quasars, and BL Lacs, have maximum
X-ray-to-optical flux ratios corresponding to log $(f_X/f_{opt})$
values of about $-1$, $0$, $+1$, and $+1.5$
\citep[e.g.,][]{stocke91,zickgraf03}. Given the typical ChaMP $0.5-2$~keV flux\footnote{This flux is the peak of an $f_X$ histogram of ChaMP sources and corresponds approximately to a $50\%$ completeness limit across the survey.}, $f_X = 10^{-14}$ erg cm$^{-2}$
s$^{-1}$, this implies that an optical counterpart for each of these
categories of typical X-ray sources will be brighter than $19, 21,
24$, and $25$ mag, respectively. As a result, all but the very
faintest stellar optical counterparts to ChaMP sources are bright enough to
have confident SDSS photometric detections. Furthermore, such targets
may be targeted for SDSS spectroscopy, allowing for secure identifications. 
 
\section{Identifying Candidate Stellar Sources}\label{id}

\subsection{Matching To SDSS}\label{match}

We begin by searching the ChaMP catalog for sources with SDSS counterparts within $20$\arcsec\ of each X-ray source centroid. We identify all potential SDSS matches to a ChaMP source and we record their distance from the X-ray centroid, along with a ratio of that distance to a radius characterizing the $95\%$ X-ray position error. The latter depends on both the number of X-ray source counts and the {\it Chandra} off-axis angle \citep{DKim04a}. We then inspect each X-ray source on the smoothed {\it Chandra} X-ray image and flag potentially contaminated sources, e.g. those that lie in the outskirts of bright X-ray sources. Detections that appear to be X-ray artifacts are also flagged, but not removed at this stage (see \S\ref{x_cuts}). Using the SDSS Image Tool \citep{sdss_images}, we simultaneously create SDSS finders for each possible optical match to the X-ray source. Here again, contaminants and potential artifacts (saturation spikes, chip edges, high background regions, etc.) are noted.

During this visual inspection, a confidence rating is attached to each match from 0 to 3, with 3 being the highest confidence match. While we flag optically saturated objects during visual inspection, these are not rejected. A match confidence of 3 typically represents a single optical counterpart with a positional offset (X-ray to optical) no greater than $2$\arcsec\ and/or less than the $95\%$ X-ray position error.

We restrict our analysis here to ChaMP sources with a match confidence of 3 and SDSS counterparts with $r<20.5$, a conservative estimate of the faintest magnitude for which SDSS performs robust morphological star/galaxy separation (see \S\ref{star_gal_sep}) even under poor observing conditions \citep{Scranton2002}. The resulting catalog contains $2121$ ChaMP sources, of which $1320$ are classified by SDSS as point sources.

\subsubsection{Estimating The Fraction Of Spurious SDSS Matches}\label{test_spurious}

\begin{figure}
\plotone{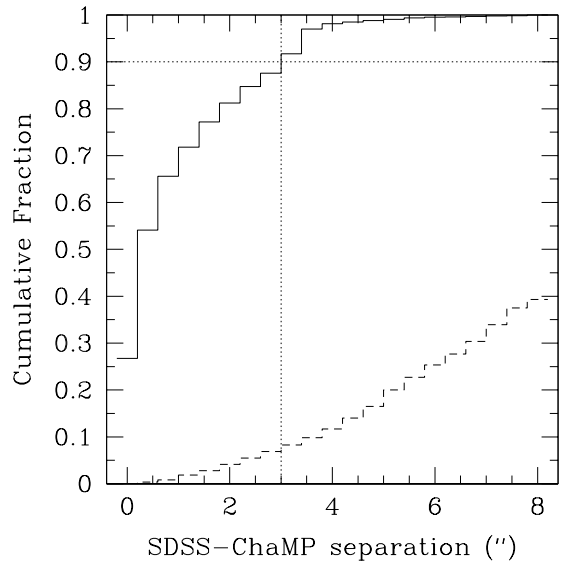}
\caption {\normalsize{{\it Solid line:} Cumulative distribution of
separations between X-ray and optical counterparts for real ChaMP/SDSS
sources with $r<20.5$ mag. {\it Dashed line:} Distribution of
separations returned by matching shifted X-ray sources to catalog of
SDSS objects with $r< 20.5$.}}\label{matching}
\end{figure}

We calculated the separation between the X-ray and optical
positions of the $2121$ matched objects selected in \S\ref{match}, 
finding a median X-ray/optical separation of $0.37$\arcsec\, with $\sigma =
1.34$\arcsec. In Figure~\ref{matching} we show the normalized
cumulative histogram of these separations; $90\%$ of the matched
sources have positions in the X-ray and optical catalogs within
$3$\arcsec\ of each other.

We then shifted the X-ray source declinations by $+30$\arcsec\
and searched for SDSS matches with $r < 20.5$ within $8$\arcsec\ of
these new positions, since only one of our original $2121$ matched
objects have separations larger than this. This procedure yields a
control sample of $833$ matches to these offset X-ray positions.

Figure~\ref{matching} shows the (dashed) cumulative normalized
histogram for this control sample; as expected, the cumulative
fraction rises with separation. Note that the normalization used here
is also $2121$, so that the dashed histogram shows an upper limit to
the fractional contamination of our sample by chance superpositions of
independent X-ray and optical sources. At $3$\arcsec, the
contamination is about $7\%$. At $4$\arcsec, an X-ray/optical
separation larger than or equal to that for $99\%$ of our sources, the
contamination is about $12\%$. This represents a conservative upper limit, since
no SDSS cuts other than $r<20.5$ have been made.

\subsection{Matching To 2MASS}

The Two Micron All Sky Survey (2MASS) obtained near-infrared images of $99.998\%$ of the sky between 1997 and 2001 \citep{Skrutskie1997,Cutri2003,Skrutskie2006}. The limiting (Vega-based) magnitudes for $10\sigma$ detections of point sources correspond roughly to $J=15.8$, $H=15.1$,  and $K_s=14.3$ mag. Positional uncertainties are $<0.2$\arcsec.

We used the Gator interface\footnote{\tt
http://irsa.ipac.caltech.edu/applications/Gator/} to identify 2MASS
counterparts for objects in our catalog, using a $3$\arcsec\ matching
radius centered on the X-ray/optical source's SDSS position. For
objects with multiple 2MASS sources within $3$\arcsec, only the
closest match was retained. This identified 2MASS counterparts for
$889$ of the $2121$ objects in our initial catalog. We also performed
a test similar to that described in \S\ref{test_spurious} to estimate
the likelihood of spurious SDSS/2MASS matches by applying a $30$\arcsec\ offset to 
each source's SDSS position and then identifying 2MASS counterparts 
within $10$\arcsec. These false matches tend to 
have SDSS/2MASS separations of $7-9$\arcsec, with $90\%$ lying outside 
of $3$\arcsec. The real matches, on the other hand, are all within 
$3$\arcsec; $97\%$ are within $1$\arcsec.

\section{Confirming The Stellar Sources}\label{pure}

\subsection{ChaMP Spectroscopy}

We queried the ChaMP spectroscopic database for existing observations
and/or classifications of objects in our catalog. All of the spectra
in the ChaMP database have been inspected and visually classified by
members of the ChaMP collaboration as either AGN/QSOs, galaxies, or
stars. $773$ sources in our sample have high confidence
classifications in the ChaMP spectroscopic database: of these, $92$
have been classified as stellar sources,
with the remaining $681$ classified as extragalactic and possessing redshifts measured using
the IRAF task {\it xcsao}. These spectral classifications
informed the criteria we develop to remove non-stellar contamination
from our sample.

\subsection{SDSS Star/Galaxy Separation}\label{star_gal_sep}

While SDSS provides automated morphological information for all
objects it detects, many of the X-ray sources in our sample have
optical counterparts significantly brighter than the SDSS saturation
limit ($\sim15$ mag). The image flux distribution of saturated stars
deviates strongly from a standard PSF and saturated stars are often
classified as extended objects. To ensure accurate morphological
classifications, we visually classified the $503$ objects with
$r<18$. We identified $53$ saturated stars misclassified as extended
sources by the SDSS pipeline, and we adjusted their entries in our
catalog.

We also checked the accuracy of the automated SDSS morphological
classification by comparing the spectroscopic and photometric
classifications of the $298$ morphologically extended objects in our
catalog with ChaMP spectra. All but five are classified
spectroscopically as extragalactic: $115$ are classified as galaxies
and $176$ as AGN/QSOs. Visual inspection of the SDSS images of these
five objects reveals that three (CXOMP J143819.2$+$033349,
J112740.4$+$565309, and J113311.9$+$010017) are extended galaxies,
suggesting their spectroscopic classification as stars is
erroneous. By contrast, CXOMP J142429.9$+$225641 and
J235645.8$-$010138 are likely stars: they are only
marginally resolved and may be either visual binaries or objects with
photometric flaws resulting in morphological misclassification.

Of the $298$ optically extended objects for which we have spectra, therefore,
only two appear to be misclassified stars based on their
photometry. This implies that $\lesssim0.7\%$ of the objects
classified as extended by the SDSS photometric pipeline are actually
point sources. Given this, we exclude from further analysis the $748$
sources whose optical counterpart has been identified as extended by
the pipeline. This increase in sample purity comes at the cost of
excluding $\sim$ five real point sources from our sample,
which does not significantly affect our completeness.

\begin{figure*}
\epsscale{0.95}
\plotone{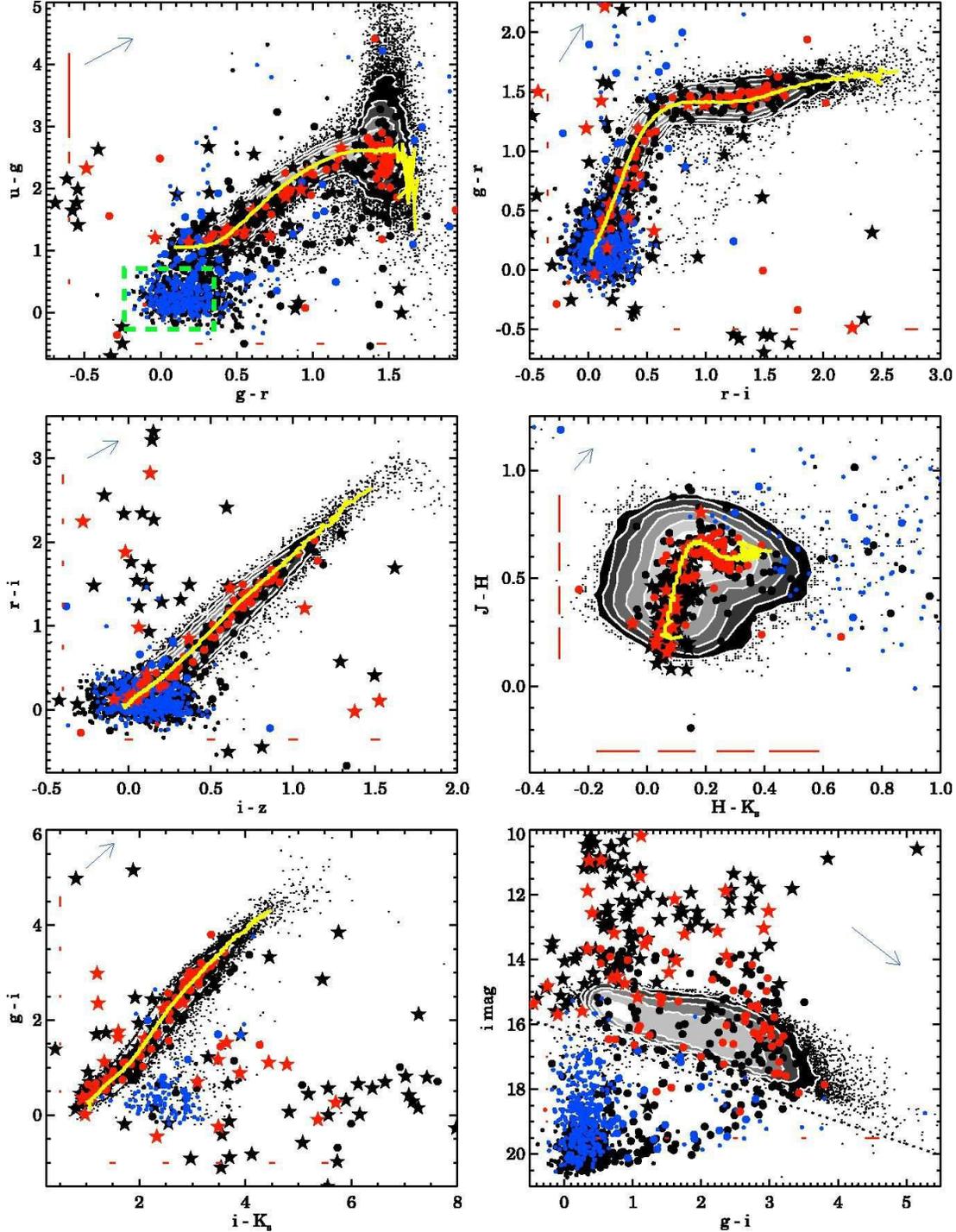}
\caption {\normalsize{The location of our initial catalog in
color-color and color-magnitude spaces. All $1373$ ChaMP/SDSS point
sources are shown as filled symbols, with stars and circles indicating
saturated and unsaturated counterparts respectively. The $87$
spectroscopically identified stars are red, while the $388$
extragalactic sources are blue. Objects in the DR5 QSO
catalog are shown with half-sized symbols; the green box in the upper
left panel is the area of color space typically inhabited by $z < 2.5$
QSOs. Grayscale contours and black dots show the high quality sample
of SDSS/2MASS point sources presented by \citet{Covey2007}; the yellow
line is the median color-color relation of this sample. The color-magnitude cut described 
in \S\ref{useCMDcut} to eliminate QSOs is shown as a dotted line in the $i$ vs.\ $g-i$
CMD. Extinction vectors corresponding to A$_V=1$ are shown with a blue arrow in the 
upper left corner of each color-color diagram, and in the upper right 
of the color-magnitude diagram.  The red bars along each axis represent 
the typical photometric errors. }}\label{fullcatalog}
\end{figure*}

Figure~\ref{fullcatalog} presents the $1373$ point sources in our
initial catalog in various optical and infrared color-color and
color-magnitude spaces. $475$ of these point sources have
spectroscopic classifications; $87$ are identified as stars and $388$
as extragalactic in nature. We highlight these two spectroscopic
samples in Figure~\ref{fullcatalog}.

\subsection{The SDSS Photometric QSO Catalog}

The SDSS provides the largest, most uniform sample of photometrically
selected quasars to $i < 21$, assembled using a nonparametric Bayesian
classification based on kernel density estimation \citep{richards04,
richards06, richards07}. Each object in the catalog is assigned a
photometric redshift according to the empirical algorithm described by
\citet{weinstein04}; the difference between the measured color and the
median colors of quasars as a function of redshift is minimized. The
quasar catalog utilized in this work includes $\sim10,000$ SDSS Data
Release 5 \citep{Adelman-McCarthy2007} photometrically selected QSOs
that fall within $20$\arcmin\ of a ChaMP field center
(G.\ Richards, private communication, 2006; Green et al.\ 2008, in
preparation). To minimize QSO contamination, we eliminate from consideration
the $827$ candidate stellar X-ray sources that are listed in the DR5 QSO catalog.

\subsection{A Color-Magnitude Cut}\label{useCMDcut}

While matching to the photometrically selected DR5 QSO catalog excludes the vast majority of QSOs in our sample, $47$ of the remaining $546$ stellar candidates are identified as QSOs in the ChaMP spectroscopic database. As the $g-i$ vs. $i$
color-magnitude diagram (CMD) in Figure~\ref{fullcatalog} shows, these
QSOs are significantly fainter ($\geq 2$ mag) than spectroscopically
confirmed stars with similar $g-i$ colors. This suggests that a
color-magnitude cut can be used to separate stars from QSOs. However,
$175$ objects still under consideration at this stage are bright
enough to saturate pixels in one or more of the five SDSS images, and
their SDSS-based colors are untrustworthy.

We therefore restrict our final sample to the $363$ sources whose
optical counterparts are either flagged as SATURATED in the SDSS
database \citep[for a detailed discussion of the SDSS flags,
see][]{stoughton02} or are unsaturated and satisfy the $i < 16.2+0.7 \times (g-i)$ color-magnitude cut shown in
Figure~\ref{fullcatalog}.  Visual inspection confirms that the $27$
objects that are saturated and do not meet our color-magnitude cut are
in fact stars.

\subsection{X-ray Quality Cuts}\label{x_cuts}

We now examine the X-ray properties of the $363$ remaining ChaMP sources to identify potential contaminants.

\begin{itemize}
\item $27$ sources are more than $12\arcmin$ from the {\it Chandra} optical axis and are subject to larger photometric and astrometric errors. Since almost all have a large number of counts, we preserve them in our sample.  We do flag these sources in our final catalog, however, and we conservatively increase their X-ray flux errors by $50\%$. 

\item $16$ sources are detected on ACIS S4, which suffers from increased noise and streaking relative to the other {\it Chandra} CCDs. These sources are flagged in our final catalog; we conservatively increase their X-ray flux errors by $20\%$.

\item We find that $14$ sources overlap according to the criteria of \citet{MKim07a}.  For eight, the overlap is small \citep[as defined by][]{MKim07a} and the X-ray photometry is reliable. For the other six, the overlap is large: we flag these sources in our catalog and conservatively double their X-ray flux errors.

\item The exposure times for nine sources are typically less than half the maximum exposure time for their respective CCDs, indicating that the source extraction region encompasses an edge or gap. These sources have unreliable fluxes and we remove them from our sample. 

\item We checked a time-ordered list of photons inside the extraction region for each source in our catalog. We searched for two consecutive photons for which the chip coordinates are the same or differ by one pixel, the exposure frames (typically $3.2$~s) increase by $1$ or $2$, and the energies decrease monotonically; these are features associated with cosmic ray afterglows\footnote{For a description of this problem, see {\tt http://asc.harvard.edu/ciao/caveats/acis\_caveats\_071213.html}.}. We remove the three false sources (all with $<10$ counts) we found in this manner from our catalog.
\end{itemize}

In summary, we remove $12$ sources from our catalog based on their X-ray properties.

\begin{deluxetable*}{lccc}
\tablewidth{0pt}
\tabletypesize{\scriptsize}
\tablecaption{Stages in catalog construction. \label{cull}}
\tablehead{
\colhead{} & \colhead{Total}   & \multicolumn{2}{c}{Spectroscopic}   \\
\cline{3-4}
\colhead{} & \colhead{Objects} & \colhead{Stars} & \colhead{Galaxies}
}
\startdata
Matched ChaMP/SDSS catalog & $2121$ & $89$ $(100\%)$ & $684$ $(100\%)$ \\
Matched ChaMP/SDSS point sources & $1373$ & $87$ $(98\%)$ & $388$ $(57\%)$ \\
... not in DR5 QSO catalog & $546$ & $86$ $(97\%)$ & $47$ $(7\%)$ \\
... with $i < 16.2+ 0.7\times(g-i)$ & $363$\tablenotemark{a} & $82$ $(92\%)$ & $3$ $(<0.1\%)$ \\
... with clean X-ray properties & $351$ & $ 81$ $(91\%)$ & $3$ $(<0.1\%)$ \\
Final catalog & $348$\tablenotemark{b} & $81$ $(91\%)$ & $0$ $(0\%)$ \\
\enddata
\tablenotetext{1}{Includes $27$ saturated stars that do not meet this color-magnitude cut.}
\tablenotetext{2}{Three spectroscopically confirmed QSOs, and 11 sources with sub-standard X-ray detections are removed manually.}
\tablecomments{Columns 3 and 4 give the number of spectroscopically confirmed stars and galaxies present in the catalog at each stage. The numbers in parentheses correspond to the fraction of the original number of these objects that is retained.} 
\end{deluxetable*}

\section{The ChaMP/SDSS Stellar Catalog: ChESS}\label{science}

Imposing the criteria described above on our initial catalog of $2121$
ChaMP detections results in a high confidence sample of $351$ stellar
X-ray emitters. This sample excludes $99.6\%$ ($681/684$) of the
spectroscopically identified extragalactic objects and includes $91\%$
($81/89$) of the spectroscopically identified stars. Of the eight
spectroscopic stars eliminated from our sample, two lack SDSS
counterparts with point source morphology, 
one is erroneously listed as having a photometric $z$ in
the SDSS QSO catalog, four fail to meet our color-magnitude
cut, and one has an X-ray detection on the edge of a {\it Chandra} CCD. We discuss the six eliminated stars with point source 
SDSS counterparts in \S\ref{stellar-sample}.

We remove the three remaining spectroscopically identified QSOs from
our sample to produce a final catalog of $348$ stellar X-ray emitters, which 
we define as the ChaMP Extended Stellar Survey
(see Table~\ref{cull} for a summary of the stages in the catalog
construction). The $348$ ChESS stars represent $17\%$ of the ChaMP sources
with SDSS counterparts, a fraction consistent with that found by 
\citet{lopez07}, as expected. X-ray and optical/near-infrared properties of 
objects in this catalog are presented in Tables~\ref{tab:ChaMPstars-xrays} and
\ref{tab:ChaMPstars-oir}.

\subsection{Previously Cataloged Stars} \label{simbad}

A number of ChESS stars are optically bright enough to have been
previously cataloged. We search for entries in the SIMBAD catalog
within $10$\arcsec\ of the ChESS position for the $348$ stars 
and find that $89$ have matches. These stars are discussed in 
more detail in Appendix~\ref{ap_1}.

The $89$ stars can be divided into three groups. The largest group, $66$ stars, 
is made up of optically bright stars that have yet to be identified as X-ray
emitters. The first group's natural complement is the small number of
stars that have already been identified as X-ray sources; there are
only $10$ stars for which this is the case. The third group is of
ChESS sources included in previous X-ray catalogs but not yet
identified; there are $13$ such sources. The vast majority of the
objects in our catalog, therefore, represent new stellar
identifications: previously known stellar X-ray sources make up
$< 3\%$ of our sample. 

\subsection{Spectroscopic Stellar Sample} \label{stellar-sample}

We used the Hammer \citep{Covey2007}, an Interactive Data Language
code\footnote{Available from {\tt
http://www.cfa.harvard.edu/$\sim$kcovey/}.} to obtain spectral types
for the $81$ stars in our sample for which we have spectra. The Hammer
predicts the Morgan-Keenan (for stars earlier than M) or Kirkpatrick
(for later stars) spectral type for a given star on the basis of a fit
to a set of $30$ spectral indices. In addition, the user can
interactively modify the assigned spectral type. Employing this tool
every spectrum was checked by eye and stars were assigned types
independently by two authors (MAA, KRC). Cases where the types
disagreed by more than two subclasses were reexamined. The spectral
types ultimately assigned are in Table~\ref{tab:ChaMPstars-specs}.

\begin{figure}
\plotone{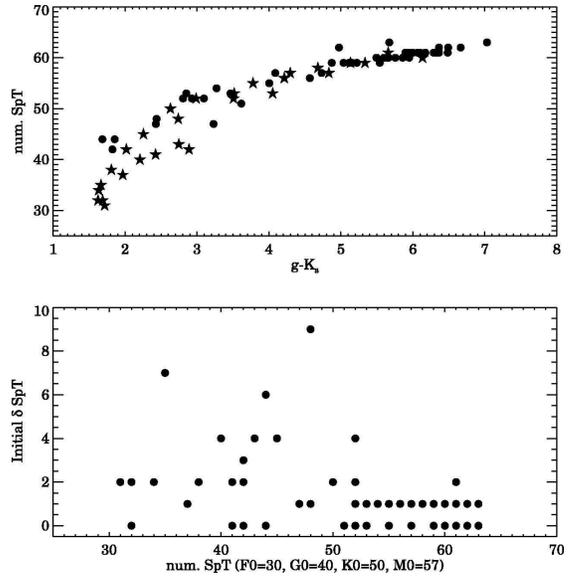}
\caption {\normalsize{{\it Top panel:} Assigned spectral types as a
function of $g-K_s$; saturated and unsaturated sources are shown as
stars and circles respectively. {\it Bottom panel:} Initial spectral
type uncertainty as a function of assigned type.}}\label{sptfig}
\end{figure}

The top panel of Figure \ref{sptfig} shows the relationship between
the assigned spectral types and each star's $g-K_s$ color; the close
relation between the two quantities (especially for unsaturated stars)
suggests that the assigned types are accurate. As an additional test
of this accuracy, we plot in the bottom panel of Figure \ref{sptfig}
the difference between the two types initially assigned to each
star. The mean difference is slightly more than one subclass, although
the quality of the agreement is dependent on the spectral type of the
star. The initial independent classifications for K and M class stars
typically disagree by one subclass or less, while initial
classifications for earlier F and G class stars typically disagree by
$2-4$ subclasses. We note that while eight of these stars have SIMBAD
entries, only three have previously cataloged spectral types and only
one is a previously known X-ray emitter. We identify CXOMP
J025951.7$+$004619 as [BHR2005] 832$-$7, which we classify as an M5
star and which SIMBAD lists as an M5.5V star. CXOMP J122837.1$+$015720
is the known X-ray emitter GSC 00282$-$00187, classified as an M2
star; we have it as an M1 star. Finally, we identify CXOMP
J231820.3$+$003129 as the F2 star TYC 577$-$673$-$1; SIMBAD lists this
star as an F5.

We list H$\alpha$ equivalent widths (EqWs) for each star
in Table~\ref{tab:ChaMPstars-specs}, which we measure by dividing 
the line flux within a $20$ \AA\ window centered at $6563$
\AA\ with the continuum flux level determined from a linear fit to two
regions ($6503-6543$ \AA\ and $6583-6623$ \AA). We then use
the $\chi$ factor \citep{Walkowicz2004} to calculate L$_{\rm
H\alpha}$/L$_{\rm bol}$ from these EqWs for the M stars with H$\alpha$
emission.  

As mentioned above (\S\ref{star_gal_sep}), the cuts we use to identify
a high confidence sample of stellar X-ray sources remove five
spectroscopically confirmed stars from our catalog. 
CXOMP J114119.9$+$661006 and J234828.4$+$005406 are optically
faint main sequence stars with spectral types K7 and M2 and are
eliminated by our color-magnitude cut; we remove an M2 star, CXOMP J161958.8$+$292321, 
because its X-ray detection falls on the edge of a {\it Chandra} CCD. The remaining three sources are
rarer cataclysmic variables, which frequently share color space with
QSOs:

\begin{itemize}
\item SDSS J020052.2$-$092431 is a previously unknown CV. Follow-up
optical observations are required to determine the nature of the
system and its period. Its soft ($0.5-2.0$ keV) flux is $3.13\pm0.28
\times 10^{-14}$ ergs cm$^{-2}$ s$^{-1}$, while its broadband ($0.3 -
8.0$ keV) flux is $9.04\pm0.65 \times 10^{-14}$ ergs cm$^{-2}$
s$^{-1}$. This CV is eliminated by our color-magnitude cut.

\item SDSS J150722.33$+$523039.8 was identified as a CV by
\citet{paula4}. Follow-up photometry revealed that it is an eclipsing
system with an extremely short orbital period of only $67$
minutes. Furthermore, observations of systems with similarly broad
absorption in the Balmer lines suggest that this CV may contain a
pulsating WD \citep[e.g.,][]{woudt2004}.

An initial match to the RASS did not return an X-ray counterpart to
this CV \citep{paula4}. It was the target of a {\it Chandra} 
observation that is included in ChaMP database. The
CV's soft flux is $2.36\pm0.84 \times 10^{-14}$ erg cm$^{-2}$
s$^{-1}$, while its broadband flux is $7.33\pm1.81 \times 10^{-14}$
erg cm$^{-2}$ s$^{-1}$. This CV is listed in the SDSS QSO catalog as
having a non-zero $z$, and also is eliminated by our color-magnitude
cut.

\item SDSS J170053.29$+$400357.6 is a known X-ray emitting polar, in
which the accretion stream flows directly onto the WD's magnetic
poles, with a period of $115$ minutes \citep{paula2}. \citet{paula2}
convert RASS counts into a flux assuming that for $2$ keV
bremsstrahlung spectrum, 1 count s$^{-1}$ corresponds to a $0.1-2.4$
keV flux of about $7 \times 10^{12}$ ergs cm$^{-2}$ s$^{-1}$. In this
case, the resulting X-ray flux is $\sim4.9\times 10^{-13}$ ergs
cm$^{-2}$ s$^{-1}$. By contrast, the soft {\it Chandra} flux is
$2.07\pm0.27 \times 10^{-13}$ ergs cm$^{-2}$ s$^{-1}$, while its
broadband flux is $6.81\pm0.62 \times 10^{-13}$ ergs cm$^{-2}$
s$^{-1}$. This CV is eliminated by our color-magnitude cut.
\end{itemize}

For all three of these CVs, the broadband flux suggests there is a
hard tail to the X-ray emission.

\subsection{Giant Stars}

\begin{figure}
\plotone{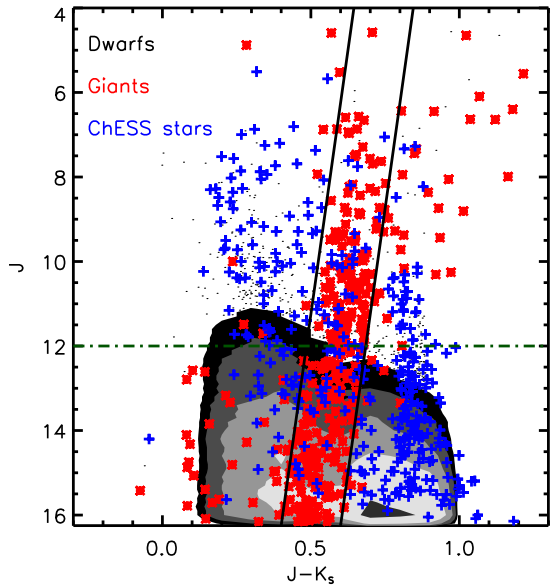}
\caption{\normalsize{Simulated $J$ vs.\ $J-K_S$ color-magnitude diagram, produced by the TRILEGAL galaxy model for SDSS/2MASS observations of a $10$ deg$^2$ field, with the ChESS stars overplotted (blue plus signs). The contours and points correspond to the distribution of $10,254$ dwarf stars (log~$g \geq 3.5$); $368$ giants are hightlighted as red asterisks. The solid lines enclose the area of the diagram in which giants are most populous. The green dot-dashed line is $J = 12$; fainter than this magnitude, giants make up only $\sim 10\%$ of the total number of stars, while brighter than this value they dominate the stellar population. We estimate that $\sim 10\%$ of the ChESS stars are giants.}}
\label{trilegalcompare}
\end{figure}

In order to estimate the fraction of ChESS stars that is likely
to be made up of evolved X-ray emitters, we generate simulated SDSS/2MASS
observations using the TRILEGAL code \citep{Girardi2005} and standard 
Galactic parameters. In Figure~\ref{trilegalcompare} we show the resulting 
$J$ vs.\ $J-K_S$ CMD. Dwarf stars are defined as having surface gravities 
log~$g \geq 3.5$ and their distribution is shown by the density contours 
and points. The positions of the simulated giant stars are given by red 
asterisks. TRILEGAL predicts that most giants ($78\%$) should reside in a fairly 
narrow locus in $J$ vs.\ $J-K_S$ color-magnitude space that stretches
from $J\sim4$ and $0.625 \leq J - K_S \leq 0.825$ down to 
$J\sim16$ and $0.4 \leq J-K_S \leq 0.6$; we highlight this region of the CMD.
We then plot the positions of the ChESS stars; $57$ inhabit the giant region. 
However, the relative fraction of giants is not uniform across this 
region. For stars with $J > 12$ mag, giants represent no more than $11\%$ of 
our simulated SDSS/2MASS detections, while they dominate the simulated stellar 
population at brighter magnitudes. Naively we would therefore only 
expect $3$ of the $29$ ChESS stars in the giant region with $J > 12$ to be 
giants; conversely, all $28$ ChESS $J < 12$ stars in this region are strong
giant candidates. Overall, this implies that $\sim 10\%$ of our sample
is made up of giant stars. Our matching to SIMBAD, discussed in \S\ref{simbad}, 
identified five known luminosity class III and IV counterparts to ChaMP sources,
as well as an RR Lyrae and a candidate Cepheid (see Appendix~\ref{ap_1}), 
implying that the minimum fraction of ChESS giants is $2\%$.

\subsection{Stellar Distances}

We wish to derive distances for the ChESS stars using photometric
parallax relations appropriate for dwarfs on the main
sequence, since these dominate our sample. However, distance estimates 
based on SDSS photometry are
unreliable for the $175$ saturated stars in our sample. Fortunately,
the SDSS photometric pipeline identifies each object's counterpart in
the USNO-B catalog \citep{monet2003}; similarly, 2MASS uses a
$5$\arcsec\ matching radius to identify counterparts in the Tycho 2 or
UNSO-A2.0 catalogs. As a result, we have either USNO or Tycho
counterparts for $347$ of the $348$ stars in our sample.

We use the Tycho/USNO $B$ magnitudes to construct $B-K_s$ colors for
each source in the catalog and derive a relationship between $g-K_s$
and $B-K_s$ for the unsaturated stars:
\begin{equation}\label{syn}
g-K_s = 0.93 \times (B-K_s)+0.25.
\end{equation} 

Comparisons of the synthetic $g-K_s$ obtained using Equation~\ref{syn}
to the measured $g-K_s$ for the unsaturated stars reveals that the
synthetic $g-K_s$ color is accurate to within $0.3$ mag ($1 \sigma$), which we
adopt as the characteristic uncertainty for our synthetic $g-K_s$.

We then generate synthetic $g-K_s$ for the $165$ saturated SDSS stars
with $B$ magnitudes. We include in Table \ref{tab:ChaMPstars-oir} the
synthetic $g$ predicted for each star (calculated from its
synthetic $g-K_s$ and the observed $K_s$), as well as a saturation
flag that indicates if a star is unsaturated, saturated in SDSS with a
synthetic $g$ from Tycho/USNO photometry, or saturated in SDSS and
lacking a Tycho/USNO counterpart.

Finally, we use a preliminary fit to the M$_{K_s}$ vs.\ $g-K_s$ CMD of
Golimowski et al.\ (2008, in preparation), which agrees well with the
tabulations of \citet{Kraus2007}, to derive distances to each star,
using synthetic $g-K_s$ colors for stars with saturated SDSS
photometry when possible. One star in our sample, CXOMP
J153203.5$+$240501, is undetected in 2MASS, so we estimate its
distance using a preliminary fit to the M$_i$ vs.\ $g-i$ CMD of
Golimowski et al.\ (2008).

\begin{figure}
\plotone{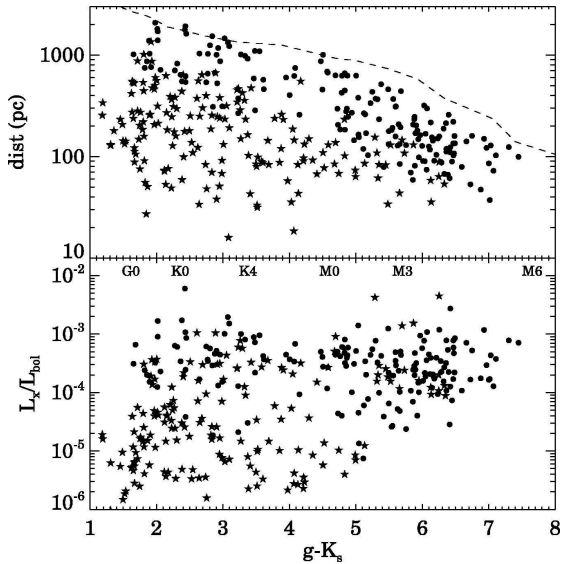}
\caption {\normalsize{{\it Top Panel:} The distance to ChESS stars as a
function of $g-K_s$ color. Stars with unsaturated SDSS photometry and
clean X-ray detections are shown as points; those with
saturated SDSS photometry and/or flagged X-ray detections 
are shown as stars. The dashed line is the
distance limit imposed by the $i$ vs.\ $g-i$ CMD cut described in
\S\ref{useCMDcut}. {\it Bottom Panel:} L$_{\rm X}$/L$_{\rm bol}$ as a
function of $g-K_s$.}}\label{gkvsdist}
\end{figure}

The resulting distances are shown in Figure \ref{gkvsdist} as a
function of $g-K_s$; formal uncertainties in these distances are $<10\%$, 
but we adopt conservative uncertainties of $20\%$ to account for 
potential systematic errors in the underlying parallax relations.  
An estimate of the distance limit imposed by the
$i$ vs.\ $g-i$ cut described in \S\ref{useCMDcut}, calculated as a
function of $g-K_s$ via the color-magnitude data tabulated by
\citet{Kraus2007}, is shown in Figure \ref{gkvsdist} as a dashed
line. This limit matches the observed upper envelope of the ChESS
catalog well. The optical/near-infrared CMD cut imposes
implicit distance limits of between $2000$ and $1000$ pc for G
and K stars and of $1000$ to $200$ pc for stars with spectral types M0 to M6. 
 
Five stars in the ChESS catalog have formal distance estimates
placing them within $20$ pc; all five have SIMBAD counterparts.
Two, CXOMP J080500.8$+$103001 and J144232.8$+$011710, are identified
as giant stars, rendering our main sequence distance estimates
invalid. Two others, CXOMP J171954.1$+$263003 and J171952.9$+$263003,
appear to be members of a binary system, despite rather different
photometric distance estimates ($8.2$ and $5$ pc); a
trigonometric parallax has been derived for the brighter component
(J171954.1$+$263003/V647 Her), placing the system at a distance of
$12$ pc. The last of the five, CXOMP J080813.5$+$210608/LHS 5134, is
also likely to be nearby: it is identified in SIMBAD as an M2.5 star,
with a distance estimate of $\sim$10 pc from spectroscopic parallax.

\subsection{Stellar X-ray Luminosities} \label{sec:lx}

\begin{figure}
\plotone{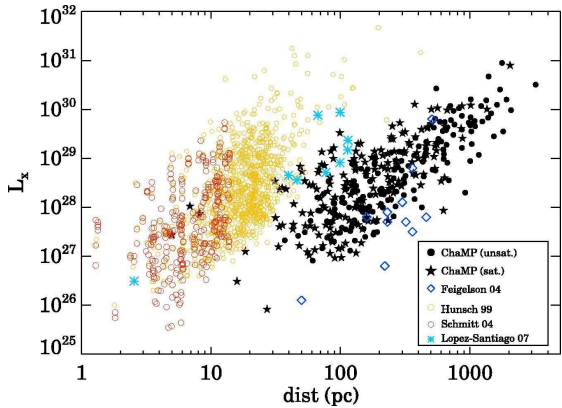}
\caption {\normalsize{L$_{\rm X}$ as a function of distance for 
several samples of X-ray emitting stars. ChESS stars with unsaturated 
SDSS photometry and clean X-ray detections are shown as filled circles; 
those with saturated SDSS photometry and/or flagged X-ray detections 
are shown as stars. Also shown are the samples of
\citet{schmitt2004} (red circles), \citet{Hunsch1999} (yellow circles),
\citet{feigelson04} (blue diamonds), and \citet{lopez07} (cyan
asterisks).}}\label{lxvsdist}
\end{figure}

Having estimated the distances to our stars, we determine their X-ray
luminosities using both the soft ($0.5-2.0$ keV) and broadband
($0.5-8$ keV) ChaMP fluxes, whose construction is described in 
\citet{MKim07a}\footnote{Note that this conversion assumes a $\Gamma = 1.7$ 
power-law X-ray spectrum; variations in coronal temperature and
metallicity can produce count to flux conversion factors that differ
by a factor of two.}. The resultant L$_{\rm X}$ values are
included in Table~\ref{tab:ChaMPstars-xrays}; here 
we limit our discussion to soft X-ray luminosities for comparison purposes. 
These luminosities are shown in Figure~\ref{lxvsdist}
as a function of distance, along with data from
several other catalogs of stellar X-ray emitters.
The primary source of the comparison data presented here
is {\it ROSAT}: we include the \citet{schmitt2004} and
\citet{Hunsch1999} catalogs ($0.1 - 2.4$ keV luminosities). We also
include the $11$ stars identified by \citet{feigelson04} in the CDF-N
($0.5-2$ keV) and the nine stars in the \citet{lopez07} {\it XMM} BSS
sample ($0.5-4.5$ keV) for which they provide distances. Compared to
these surveys, the ChESS catalog samples a unique area in the
L$_{\rm X}$--distance plane, covering the ranges of
$2\times10^{26}\ \lapprox\ $L$_{\rm X}\ \lapprox\ 2\times10^{31}$ ergs s$^{-1}$ and $30\ \lapprox\ $d $\lapprox\ 3000$ pc.

The ChESS stars are for the most part more luminous than those
in the volume complete sample assembled by \citet{schmitt2004}. 
Despite their low intrinsic luminosities, the nearest stars
have moderately large X-ray fluxes ($\sim 10^{-12}$ ergs cm$^{-2}$
s$^{-1}$). Fields in the {\it Chandra} archive including such sources
are explicitly excluded from the ChaMP survey: the increased
likelihood of saturation in X-ray and optical imaging reduces the
ability to detect and classify other X-ray sources in the field, and
greatly complicates the calculation of the effective area sampled by
the observation.

The larger catalog of stellar X-ray emitters assembled by
\citet{Hunsch1999} provides a more natural comparison to our ChESS
catalog. The L$_{\rm X}$ lower limit of each sample increases with
distance, as expected for flux-limited catalogs. While the distance
limit of the ChESS catalog is fundamentally optical in nature
(due to the CMD cut described in \S\ref{useCMDcut}), a crude
comparison of the relative sensitivities of the surveys 
can be made by comparing the distances to which each
instrument can detect stars of a given L$_{\rm X}$: the
\citet{Hunsch1999} sample includes stars with L$_{\rm X} = 10^{28}$
ergs s$^{-1}$ to a distance of $30$ pc, while the ChESS catalog
contains such stars out to $200$ pc. The surface density of stars
in the ChESS catalog ($\sim 10$ deg$^{-2}$) exceeds that of
the \citet{Hunsch1999} catalog ($3\times10^{-4}$ deg$^{-2}$) by nearly
five orders of magnitude.

Figure~\ref{lxvsdist} shows that the ChESS stars'
properties are most similar to those of stars included in other {\it
Chandra} and {\it XMM} catalogs. These catalogs are not
interchangeable, however. For example, while the luminosities of the
\citet{feigelson04} CDF-N stars are comparable to those of the least
luminous members of the ChESS catalog, that sample's effective
distance limit is beyond that of the ChESS catalog for
equivalent X-ray luminosities. Conversely, because the \citet{lopez07}
sample relies on trigonometric parallax measurements for distances, these {\it
XMM}-detected stars, while also comparably X-ray luminous to the ChESS
stars, make up a shallower sample.

We also present in Table~\ref{tab:ChaMPstars-xrays} the hardness
ratio (HR) for each source, where HR $= ($H$_c-$S$_c$)/(H$_c+$S$_c$) 
and H$_c$ and S$_c$ are the number of hard and soft counts, 
respectively \citep{MKim07a}. The stars in our catalog are quite soft, 
with typical HRs from $-1.0$ to $-0.6$; HR shows no clear correlation 
with L$_{\rm X}$ or $g-K_s$.

\subsection{Stellar Bolometric Luminosities}

For each star, we derive the bolometric luminosity using the $g-K_s$ color and the appropriate \citet{Kraus2007} bolometric correction. The resulting L$_{\rm X}$/L$_{\rm bol}$ ratios are presented in Table~\ref{tab:ChaMPstars-xrays} and shown in the bottom panel of Figure~\ref{gkvsdist} as a function of $g-K_s$.

The lower limit to the L$_{\rm X}$/L$_{\rm bol}$ values in the ChESS catalog 
is shaped by the sample's effective L$_{\rm X}$ limit, 
which is a function of the exposure times of the {\it
Chandra} images used to build the ChaMP.  The presence of an upper
envelope at L$_{\rm X}$/L$_{\rm bol} \sim 10^{-3}$, however, reflects
a physical characteristic of the stars.  Previous investigators have
found a similar empirical upper limit to the efficiency of stellar
X-ray emission \citep[e.g.,][]{Vilhu1983,Vilhu1987,Herbst1989,Stauffer1994}.  
While the cause of this so-called saturation is still unknown, it
is most commonly attributed to feedback processes that quench the
efficiency of the stellar dynamo and/or the ability of the dynamo to
heat the coronal plasma \citep{Cameron1994}, or to centrifugal
stripping of the coronal plasma at the high rotational velocities
associated with large L$_{\rm X}$ \citep{Jardine2004}.

\begin{figure}
\plotone{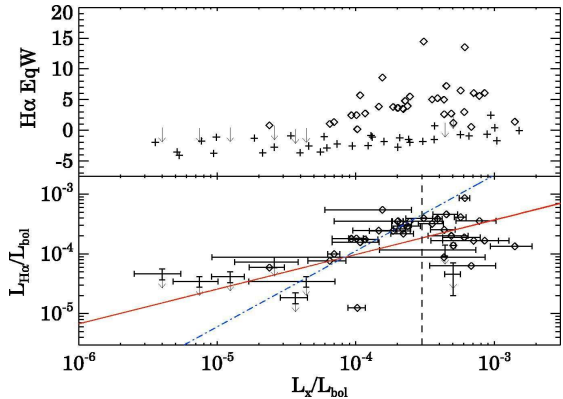}
\caption {\normalsize{ {\it Top Panel:} H$\alpha$ EqW vs.\ L$_{\rm X}$/L$_{\rm bol}$ for stars with ChaMP spectra. Negative EqWs indicate the presence of absorption lines.  F, G, and K stars are shown with plus signs; M stars are indicated with diamonds. The downward-pointing arrows indicate the EqW upper limits for M stars with no detected H$\alpha$ emission.  {\it Bottom panel:} L$_{\rm H\alpha}$/L$_{\rm bol}$ vs.\ L$_{\rm X}$/L$_{\rm bol}$ for the M stars in the spectroscopic sample, with symbols as above. The red line is the best fit relation between L$_{\rm H\alpha}$/L$_{\rm bol}$ and L$_{\rm X}$/L$_{\rm bol}$ for the entire sample. The blue-dot dashed line is the relation for the stars with L$_{\rm X}$/L$_{\rm bol} < 3\times10^{-4}$, a value indicated by the dashed line.}}\label{halxlbol}
\end{figure}

Figure~\ref{halxlbol} compares non-simultaneous measures of the 
strength of the H$\alpha$ emission line, a common diagnostic of 
chromospheric activity, with L$_{\rm X}$/L$_{\rm bol}$, a tracer
of coronal activity for stars in our spectroscopic sample.  
Similar measurements from M stars in young clusters and the solar 
neighborhood \citep[e.g., ][]{Reid1995}, have found 
L$_{\rm X} = (3-5) \times$ L$_{\rm H\alpha}$, but were typically
made using {\it ROSAT} data. 
As stellar coronae produce very soft X-ray emission, it is 
unsurprising that the ChESS data, measuring harder X-rays, produces
an L$_{\rm X}$/L$_{\rm bol}$ ratio of $\sim2/3$, lower than the {\it ROSAT}-measured
ratio by a factor of five.

The correlation between L$_{\rm X}$/L$_{\rm bol}$ and L$_{\rm H\alpha}$/L$_{\rm bol}$ in the ChESS data, however, is  highly significant by Cox Proportional Hazard ($P=0.0008$), Kendall's $\tau$ ($P=0.0027$), and Spearman's $\rho$ tests ($P=0.0064$), as implemented 
in the Astronomy Survival Analysis Package \citep{LaValley92}.
We perform bivariate linear regressions with log(L$_{\rm X}$/L$_{\rm bol}$)
as the dependent variable, using the parametric EM algorithm, 
and find the following best-fit relationship:

\begin{equation}
{\rm log( L_{H\alpha}/L_{bol}) =  (0.58\pm0.13)\times log(L_X/L_{bol}) - (1.69\pm0.48)}
\end{equation}
		
\noindent with RMS residuals of $0.39$; this relationship is shown as the red line in Figure~\ref{halxlbol}.
When restricting the sample to L$_{\rm X}$/L$_{\rm bol} <3\times 10^{-4}$, the best-fit regression line steepens to:

\begin{equation}
{\rm log(L_{ H\alpha}/L_{bol}) = (1.27\pm0.24)\times log(L_X/L_{bol}) + (1.13\pm0.94)}, 
\end{equation}

\noindent shown as the blue dot-dashed line in Figure~\ref{halxlbol}, with RMS residuals of $0.31$.

The steepening of the L$_{\rm X}$/L$_{\rm bol}$ vs.\ L$_{\rm H\alpha}$/L$_{\rm bol}$ relation when high L$_{\rm X}$/L$_{\rm bol}$ sources are excluded, and the turnover in L$_{\rm H\alpha}$/L$_{\rm bol}$ at large L$_{\rm X}$/L$_{\rm bol}$ that is clearly visible in Figure \ref{halxlbol},
reveal that stars with very active coronae can possess very pedestrian chromospheres, at least when viewed at distinct epochs. To ensure that this effect is not merely an effect of uncertain H$\alpha$ measurements in low S/N spectra, we visually inspected the H$\alpha$ region
in the stars with L$_{\rm X}$/L$_{\rm bol} > 3 \times 10^{-4}$.
We find that these spectra are of high enough quality to confirm that only very low levels of H$\alpha$ emission are present in these stars.  We also verified that there are no significant differences in the spectral type or Galactic height of stars when the sample is divided at L$_{\rm X}$/L$_{\rm bol} = 3\times$10$^{-4}$.

There exist at least two plausible explanations for this seeming disconnect between the chromospheric and coronal properties of the stars with the most active coronae:

\begin{enumerate}

\item{Our X-ray selected sample is biased towards 
detecting flaring stars, whose non-simultaneous optical spectra
may be obtained when the star has returned to quiescence.
The seeming disconnect between the coronal and chromospheric properties 
would then simply reflect the temporal disconnect in the 
observations of these stars. If this is the case, an extremely crude indicator of the duty cycle 
of X-ray flares on M stars in the Galactic disk can be derived from
the $\sim45$\% ($19/43$) of the sample with low, and presumably quiescent, 
H$\alpha$ luminosity: the failure to observe significant H$\alpha$ 
emission during spectroscopic exposures with a median length of $720$ s
would imply a upper limit on the typical flare rate of 5 H$\alpha$ flares hr$^{-1}$.}

\item{Alternatively, the lack of correlation between chromospheric and 
coronal emission may be a sign that these two types of activity
decouple as coronal activity levels approach the saturated regime. 
This hypothesis has been advanced previously \citep[e.g., ][]{Cram1982,Pettersen1987,Mathioudakis1989,Houdebine1996}; in this scenario, the relative 
efficiencies of radiative processes that cool the corona 
and chromosphere (e.g., H$\alpha$, Ca II, and Mg emission, highly 
ionized X-ray line emission, and ultraviolet continuum emission) are
sensitive to the strength of stellar activity. To explain the effect
seen here, extreme levels of stellar activity would have to quench 
cooling of the chromosphere via H$\alpha$ emission even as the 
corona continues to be cooled efficiently by X-rays.}
\end{enumerate}

The relatively weak coronae implied by the 
L$_{\rm X}/$L$_{\rm H\alpha}$ relationship measured from the 
low-activity portion of our sample and its apparent 
breakdown at high activity levels present intriguing clues to the 
temporal behavior of coronal activity over timescales characteristic of 
both the non-simultaneity effects ($t<10$ yr) and population effects ($t>1$ 
Gyr) discussed above. The current sample of stars with 
measurements of both L$_{\rm X}$/L$_{\rm bol}$ and 
L$_{\rm H\alpha}$/L$_{\rm bol}$ is too small, however, to draw firm
conclusions.  We defer a full analysis of these 
effects to follow-up studies.

\subsection{Stellar Colors}

While the clearest signatures of magnetic activity are spectroscopic
in nature, stellar activity can impact a star's broadband colors as
well.  In particular, magnetically active stars appear to possess
ultraviolet (UV) excesses of $0.03-0.1$ mag in $U-B$ compared to
non-active stars. This excess has been attributed to continuum
emission generated from hot, active chromospheres
\citep{Houdebine1996,Houdebine1997,Amado1997, James2000,Sung2002,Amado2003,Bochanski2007}.

\begin{figure}
\plotone{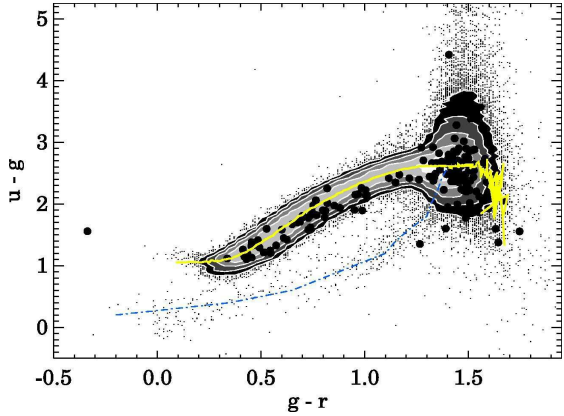}
\caption {\normalsize{$u-g$ vs.\ $g-r$ for the ChESS stars with
unsaturated SDSS photometry and unflagged X-ray detections
(dots) with the optically selected SDSS/2MASS 
sample constructed by \citet{Covey2007} 
shown for comparison, as in Figure~\ref{fullcatalog}.
The yellow line is the median stellar colors of the \citet{Covey2007}
sample; the blue dashed line shows the locus of WD/M dwarf pairs
identified by \citet{Smolcic2004}. }}\label{cleancolorcolor}
\end{figure}

The $u-g$ vs.\ $g-r$ color-color diagram in Figure~\ref{cleancolorcolor}
shows evidence for a similar shift, with X-ray emitting, optically unsaturated ChESS stars
lying systematically lower than the median $u-g$ vs.\ $g-r$ locus
measured by \citet{Covey2007} from a sample of optically selected
SDSS/2MASS stars.  This shift in color-color space, however, is not
unambiguous proof of a $u-g$ excess, as the offset could be
caused by a red excess in $g-r$, particularly since active stars can
have strong H$\alpha$ emission that contributes additional flux
to the $r$ band.  Our spectroscopic sample, however, does not include
any stars with H$\alpha$ equivalent widths significantly larger than
$10$ \AA\ (see Table~\ref{tab:ChaMPstars-specs}), and even
such strong H$\alpha$ emission lines contribute only a small
fraction to the flux transmitted through a $\sim1000$-\AA\ wide
filter, brightening a star in the $r$ band by only $0.01$ mag.

\begin{figure}
\epsscale{.8}
\plotone{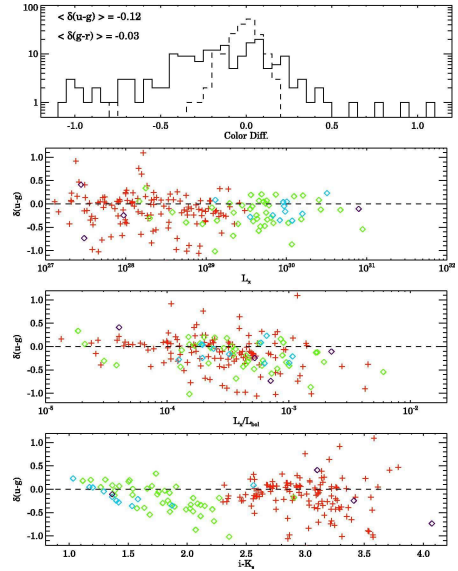}
\caption {\normalsize{ {\it Top Panel:} Histograms of color
differences between unsaturated ChESS stars and optically selected
stars with identical $i-K_s$.  Differences for $u-g$ (solid line) and
$g-r$ (dashed line) are shown. {\it Second Panel:} $u-g$ differences
for unsaturated stars as a function of L$_{\rm X}$.  M stars are
shown as red crosses, while F, G, and K stars are shown as purple, blue, and green 
diamonds respectively.  {\it Third Panel:}
$u-g$ differences for unsaturated stars as a function of L$_{\rm
X}$/L$_{\rm bol}$. {\it Bottom Panel:} $u-g$ differences for
unsaturated stars as a function of $i-K_s$.}}\label{colorshifts}
\end{figure}

To confirm that the offset in $u-g$ vs.\ $g-r$ is due to the
stars' anomalous $u-g$ colors, we compare the offsets between the $u-g$ and
$g-r$ colors of unsaturated stars in our sample and the median colors
of non-active stars with the same $i-K_s$ color tabulated by
\citet{Covey2007} (see top panel, Figure~\ref{colorshifts}). 
While the spread is large, active stars are
systematically bluer by $0.12$ mag in $u-g$ than inactive stars. By
contrast, the $g-r$ colors of active stars are consistent with those
of inactive stars to within $0.03$ mag, and there the difference is
that active stars are bluer than inactive stars. This is inconsistent
with the idea of a red shift caused by the addition of H$\alpha$
emission into a star's $r$ band.

While stellar $u-g$ colors are sensitive to metallicity
and the presence of unresolved WD companions,
neither effect is likely to explain the offset seen here.
The sensitivity of $u-g$ to metallicity is
due to line blanketing, where absorption by a large
number of metal lines in the $u$ band
leads to preferentially redder $u-g$ colors for more
metal-rich stars.  Interpreted as a metallicity
effect, however, the $\sim0.1$ mag blue $u-g$ offset 
implies that X-ray luminous stars have metallicities more than
half a dex lower than the standard field population
\citep{Karaali2005}, exceedingly unlikely given the well
known link between stellar age and X-ray luminosity.

Similarly, while main sequence stars with an
unresolved WD companion have anomalously blue
$u-g$ colors, as well as the potential for enhanced X-ray luminosity,
the colors of the stars in our sample disagree
with those expected for such binaries. The SDSS colors of 
WD/main sequence binaries found by \citet{Smolcic2004} and 
\citet{Silvestri2006} are shown in Figure~\ref{cleancolorcolor}. While there
may be a handful of such systems in our sample, the bulk of the
ChESS stars are redder in $u-g$ than would be expected for systems 
with WD components.

To investigate the cause of this $u-g$ offset,
Figure~\ref{colorshifts} also shows the magnitude of the $u-g$ offset
as a function of L$_{\rm X}$, L$_{\rm X}$/L$_{\rm bol}$, and $i-K_s$,
a proxy for stellar temperature and mass. A slight tendency for the
offset to increase with L$_{\rm X}$/L$_{\rm bol}$ may be present,
particularly when considering only stars of a given spectral type, but
linear regression does not return a statistically significant
correlation between the two variables. One would expect
the $u-g$ excess to be most prominent for M stars, which
typically have the highest activity and the lowest 
of quiescent UV flux, allowing contributions from the chromosphere to
affect the stars' $u-g$ most significantly. Instead, the $u-g$ excess
reaches a maximum for K stars (at $i-K_s \sim 2.0$) and then 
decreases into the M regime. Whether this effect is real or the result 
of observational bias is hard to access, in part because of 
the increased uncertainties in $u-g$ for late-type stars caused by the 
red leak in the SDSS camera\footnote{The red leak describes an instrumental
effect whereby the $u$-band filter transmits flux longward of $7100$
\AA\ due to changes in the filter's interference coating under
vacuum. This instrumental effect depends on a star's raw $u$ and $r$
magnitudes, which in turn are dependent on the airmass, seeing, and
the sensitivity of each $u$ filter as a function of wavelength and
stellar spectrum. Given the complexity of this effect, the SDSS
photometric pipeline does not attempt to correct each star's $u$-band
photometry, resulting in increased $u$ uncertainties of $0.02$ mag for
K stars, $0.06$ mag for M0 stars, and $0.3$ mag for stars with $r-i >
1.5$. For more information see
\url{http://www.sdss.org/dr6/products/catalogs/index.html}.}.  The
additional scatter in the $u-g$ colors of these stars may wash out
evidence for trends of $\delta(u-g)$ with either L$_{\rm X}$ or
color. Follow-up studies with more reliable $u$ photometry are 
needed to reveal the nature of any correlation between $u-g$ excess and
coronal or chromospheric activity.

\subsection{Stellar Populations} \label{thinthick}

\begin{figure}
\plotone{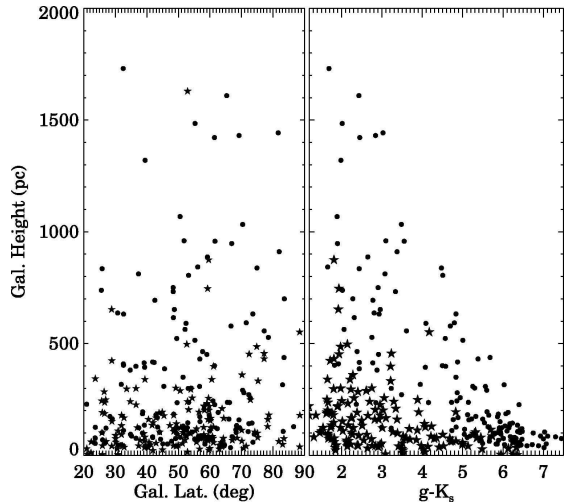}
\caption {\normalsize{{\it Left Panel:} Height in the Galactic disk
(in pc) as a function of Galactic latitude. {\it Right Panel:}
Height in the Galactic disk (in pc) as a function of
$g-K_s$.  Symbols as in Fig. \ref{lxvsdist}.}}\label{lvsz}
\end{figure}

Previous studies have found that magnetically active stars have a smaller 
Galactic scale height than non-active stars \citep[e.g., ][]{West2008}.
To determine how the stars in our catalog are distributed between the
different Galaxy components, we use each star's
distance and Galactic latitude to calculate its height in the Galactic
disk.  We show in Figure~\ref{lvsz} the resulting Galactic heights as
a function of both Galactic latitude and stellar color. If our catalog
were probing a spherically symmetric halo population, the
color-magnitude cut imposed in \S\ref{useCMDcut} would limit the
catalog mainly as a function of the heliocentric distance to each
star. Sight lines probing higher Galactic latitudes would sample stars
at larger Galactic heights.  The distribution of Galactic heights in
the sample is independent of Galactic latitude, however, indicating
that the distribution of stars within the disk of the Milky Way
imposes a stricter distance limit than the color-magnitude cut imposed
in \S\ref{useCMDcut}.

\begin{figure}
\plotone{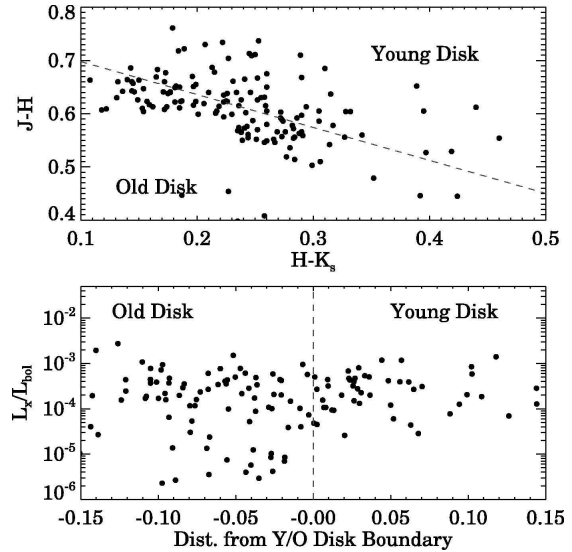}
\caption {\normalsize{{\it Top Panel: }$J-H$ vs.\ $H-K_s$ for ChESS M
stars with unflagged X-ray detections. 
The dashed line is the boundary between the regions identified
by \citet{Stauffer1986} and \citet{Leggett1992} as populated
preferentially on one side by relatively high-metallicity young disk
stars and on the other by relatively low-metallicity old disk
stars. {\it Bottom Panel: } L$_{\rm X}$/L$_{\rm bol}$ as a function of
offset in $J-H$ from the young disk/old disk boundary in the top
panel.}}\label{fehjhhk}
\end{figure}

\citet{Stauffer1986} and \citet{Leggett1992} have correlated the
near-infrared colors of M stars and their metallicities and
kinematics, allowing them to define regions of $J-H$ vs.\ $H-K_s$
color-color space dominated by young and old disk stars. In
Figure~\ref{fehjhhk}, we compare the $JHK_s$ colors of M stars in our
sample to the boundary defined by \citet{Leggett1992} between young
and old disk stars.  This boundary nearly bisects our sample,
suggesting that the ChESS catalog contains both young stars and
the high activity tail of the old disk
population. Figure~\ref{fehjhhk} also shows L$_{\rm X}$/L$_{\rm bol}$
for these M stars as a function of their offset from the
young/old disk boundary. The lowest activity sources (L$_{\rm
X}$/L$_{\rm bol} \sim 10^{-5}$) are uniformly identified with the old
disk population, a clear signature of the decay of magnetic activity
with age.  Interpreting the significance of the many old disk stars with
large L$_{\rm X}$/L$_{\rm bol}$ values is less straightforward, particularly
because these active old disk stars are likely merely color outliers 
of the vastly more numerous young disk population.  If these high 
L$_{\rm X}$/L$_{\rm bol}$ stars are truly members of the old disk, however,
they would represent a new and very significant population of stars that
experience little decay of magnetic activity over their lifetimes.

\section{Conclusions}\label{concl}

We have correlated the Extended {\it Chandra} Multiwavelength Project
with the Sloan Digital Sky Survey to identify the $348$ X-ray 
emitting stars of the ChaMP Extended Stellar Survey. We used morphological 
star/galaxy separation, matching to an SDSS quasar catalog, an
optical color-magnitude cut, and X-ray data quality tests to identify the
ChESS stars from a sample of $2121$ matched ChaMP/SDSS sources. 

\begin{itemize}
\item Our cuts retain $91\%$ of the spectroscopically confirmed stars
in the original sample while excluding $99.6\%$ of the $684$
spectroscopically confirmed extragalactic sources. Fewer than $3\%$ of
the sources in our final catalog are previously identified stellar
X-ray emitters.

\item For $42$ catalog members, spectroscopic classifications are
available in the literature. We present new spectral classifications
and H$\alpha$ measurements for an additional $79$ stars. We derive
distances to the stars in our catalog using photometric parallax
relations appropriate for dwarfs on the main sequence and calculate
their X-ray and bolometric luminosities. For $36$ newly identified
X-ray emitting M stars we also provide measurements of L$_{\rm
H\alpha}$/L$_{\rm bol}$.

\item The stars in our catalog lie in a unique space in the L$_{\rm
X}$--distance plane, filling the gap between the nearby stars
identified as counterparts to sources in the {\it ROSAT} All-Sky
Survey and the more distant stars detected in other {\it Chandra} and
{\it XMM-Newton} surveys.

\item The ChESS catalog is dominated by main sequence stars. 
By comparing the distribution of the ChESS sample in $J$ vs.\ $J-K_S$ 
space to that of simulated SDSS/2MASS observations generated by TRILEGAL, 
we estimate that the total fraction of giants in the catalog is $\sim 10\%$.
In addition to seven confirmed giant stars (including a possible Cepheid 
and an RR Lyrae star), we identify three cataclysmic variables.

\item We find that L$_{\rm H\alpha}$/L$_{\rm bol}$ 
and L$_{\rm X}$/L$_{\rm bol}$ are linearly related below 
L$_{\rm X}$/L$_{\rm bol} \sim 3 \times 10^{-4}$, while 
L$_{\rm H\alpha}$/L$_{\rm bol}$ appears to turn over at larger
L$_{\rm X}$/L$_{\rm bol}$ values.  

\item Stars with reliable SDSS photometry have an $\sim0.1$
mag blue excess in $u-g$, likely due to increased chromospheric
continuum emission. Photometric metallicity estimates suggest that our
sample is evenly split between the young and old disk populations of
the Galaxy; the lowest activity sources are identified with the old
disk population, a clear signature of the decay of magnetic activity
with age.
\end{itemize}

Future papers will present analyses of ChESS source variability and
comparisons of the ChESS catalog to models of stellar activity in the
Galactic disk.

\acknowledgements 
We thank Suzanne Hawley,  Andrew West, Steven Saar, and Thomas Fleming 
for useful discussions of stellar magnetic activity; we also thank the 
anonymous referee and editor for useful comments that improved the work 
presented here. We are indebted to the staffs at the National Optical 
Astronomy Observatories, Las Campanas, and the MMT for assistance with 
optical spectroscopy.  Special thanks to observers including Warren 
Brown, Perry Berlind, and Michael Calkins, for FAST spectroscopy from 
the Fred Lawrence Whipple Observatory $1.5$ m on Mt Hopkins, and to 
Susan Tokarz and Nathalie Marthimbeau for reductions.  

Support for this work was provided by the National
Aeronautics and Space Administration through {\it Chandra}, Award
Number AR4-5017X and AR6-7020X issued by the {\it Chandra} X-ray
Observatory Center, which is operated by the Smithsonian Astrophysical
Observatory for and on behalf of the National Aeronautics Space
Administration under contract NAS8-03060. Further NASA support was
provided to K.\ Covey through the Spitzer Space Telescope Fellowship
Program, through a contract issued by the Jet Propulsion Laboratory,
California Institute of Technology under a contract with NASA. M.\
Ag\"ueros is supported by an NSF Astronomy and Astrophysics
Postdoctoral Fellowship under award AST-0602099. D.\ Haggard is
supported by a NASA Harriett G.\ Jenkins Predoctoral Fellowship.

This work is based in part on observations obtained at Cerro Tololo 
Inter-American Observatory and Kitt Peak Observatory, National Optical 
Astronomy Observatory, operated by the Association of Universities 
for Research in Astronomy, Inc.\ under cooperative agreement with the
National Science Foundation.

This research has made use of NASA's Astrophysics Data System
Bibliographic Services, the SIMBAD database, operated at CDS,
Strasbourg, France, the NASA/IPAC Extragalactic Database, operated by
the Jet Propulsion Laboratory, California Institute of Technology,
under contract with the National Aeronautics and Space Administration,
and the VizieR database of astronomical catalogs
\citep{Ochsenbein2000}. IRAF (Image Reduction and Analysis Facility) is 
distributed by the National Optical Astronomy Observatories, which are 
operated by the Association of Universities for Research in Astronomy, 
Inc., under cooperative agreement with the National Science Foundation.

Funding for the SDSS and SDSS-II has been provided by the Alfred
P. Sloan Foundation, the Participating Institutions, the National
Science Foundation, the U.S. Department of Energy, the National
Aeronautics and Space Administration, the Japanese Monbukagakusho, the
Max Planck Society, and the Higher Education Funding Council for
England. The SDSS Web Site is {\tt http://www.sdss.org/}.

The SDSS is managed by the Astrophysical Research Consortium for the
Participating Institutions. The Participating Institutions are the
American Museum of Natural History, Astrophysical Institute Potsdam,
University of Basel, University of Cambridge, Case Western Reserve
University, University of Chicago, Drexel University, Fermilab, the
Institute for Advanced Study, the Japan Participation Group, Johns
Hopkins University, the Joint Institute for Nuclear Astrophysics, the
Kavli Institute for Particle Astrophysics and Cosmology, the Korean
Scientist Group, the Chinese Academy of Sciences (LAMOST), Los Alamos
National Laboratory, the Max-Planck-Institute for Astronomy (MPIA),
the Max-Planck-Institute for Astrophysics (MPA), New Mexico State
University, Ohio State University, University of Pittsburgh,
University of Portsmouth, Princeton University, the United States
Naval Observatory, and the University of Washington.

The Two Micron All Sky Survey was a joint project of the University of
Massachusetts and the Infrared Processing and Analysis Center
(California Institute of Technology). The University of Massachusetts
was responsible for the overall management of the project, the
observing facilities and the data acquisition. The Infrared Processing
and Analysis Center was responsible for data processing, data
distribution and data archiving.

\appendix
\section{ChaMP Sources With SIMBAD Counterparts}\label{ap_1}

In Table~\ref{tab:simbad-stars} we present the optical data for the
$66$ stars cataloged in SIMBAD that we have identified as ChaMP X-ray
sources, and include additional information (spectral type, binarity,
variability) where available. We searched the literature for evidence
that these stars had been identified as X-ray sources and could find
no previous X-ray detections; we therefore consider these all to be
new X-ray source identifications. Four stars are positionally
coincident with X-ray sources in other {\it Chandra} catalogs, but are
not explicitly listed in SIMBAD as X-ray emitters or identified in
these catalogs as stars, and we therefore consider them also to be new
identifications. CXOMP J084944.7$+$445840 is among the sources
detected in Lynx \citep{stern02} and listed in the Serendipitous
Extragalactic X-Ray Source Identification \citep[SEXSI;][]{harrison03}
catalog, but is unidentified in both catalogs. CXOMP
J085005.3$+$445819 and J090941.7$+$541939 are both unidentified SEXSI
sources. Finally, CXOMP J162157.2$+$381734 is less than $10$\arcsec\
from 1RXS J162157.6$+$381727, an unidentified RASS source.

$13$ ChaMP stellar sources do not have SIMBAD optical counterparts but
are included in other X-ray catalogs. However, our examination of
these catalogs reveals no additional information about the nature of
these sources, and we also consider these to be new X-ray source
identifications. For example, CXOMP J084854.0$+$450230 is within
$1$\arcsec\ of the X-ray source [STS2002] 43 \citep{stern02}, but the
catalog for that survey does not include an identification for this
X-ray source or for two other ChaMP sources. Similarly, eight ChaMP
sources listed in the SEXSI catalog and two observed in Bootes by
\citet{wang04} are not identified. CXOMP J141120.7$+$521411 is
included in three catalogs and unidentified in all three, although a
magnitude is given for the counterpart by \citet{zickgraf03} in their
catalog of RASS BSC sources. CXOMP J125152.2$+$000528 is listed by
\citet{zickgraf03}, but is unidentified. CXOMP
J214229.3$+$123322 is within $4$\arcsec\ of the unidentified RASS
source 1RXS J214229.5$+$123323. These sources are listed in
Table~\ref{tab:other-xray}. In total, we have $79$ ChESS stars with
cataloged optical or X-ray data, but which had not previously been
identified as stellar X-ray sources.

Finally, $10$ ChESS stars are previously known stellar X-ray
sources. We list these in Table~\ref{tab:known-sources}. A full
examination of the properties of these stars (e.g., a comparison of
their previously reported fluxes to those detected by {\it Chandra})
is beyond the scope of this paper.


\begin{thebibliography}{100}
\expandafter\ifx\csname natexlab\endcsname\relax\def\natexlab#1{#1}\fi

\bibitem[{{Adelman-McCarthy} {et~al.}(2007){Adelman-McCarthy},
  {Ag{\"u}eros}, {Allam}, {Anderson}, {Anderson}, {Annis}, {Bahcall},
  {Bailer-Jones}, {Baldry}, {Barentine}, {Beers}, {Belokurov}, {Berlind},
  {Bernardi}, {Blanton}, {Bochanski}, {Boroski}, {Bramich}, {Brewington},
  {Brinchmann}, {Brinkmann}, {Brunner}, {Budav{\'a}ri}, {Carey}, {Carliles},
  {Carr}, {Castander}, {Connolly}, {Cool}, {Cunha}, {Csabai}, {Dalcanton},
  {Doi}, {Eisenstein}, {Evans}, {Evans}, {Fan}, {Finkbeiner}, {Friedman},
  {Frieman}, {Fukugita}, {Gillespie}, {Gilmore}, {Glazebrook}, {Gray},
  {Grebel}, {Gunn}, {de Haas}, {Hall}, {Harvanek}, {Hawley}, {Hayes},
  {Heckman}, {Hendry}, {Hennessy}, {Hindsley}, {Hirata}, {Hogan}, {Hogg},
  {Holtzman}, {Ichikawa}, {Ichikawa}, {Ivezi{\'c}}, {Jester}, {Johnston},
  {Jorgensen}, {Juri{\'c}}, {Kauffmann}, {Kent}, {Kleinman}, {Knapp},
  {Kniazev}, {Kron}, {Krzesinski}, {Kuropatkin}, {Lamb}, {Lampeitl}, {Lee},
  {Leger}, {Lima}, {Lin}, {Long}, {Loveday}, {Lupton}, {Mandelbaum}, {Margon},
  {Mart{\'{\i}}nez-Delgado}, {Matsubara}, {McGehee}, {McKay}, {Meiksin},
  {Munn}, {Nakajima}, {Nash}, {Neilsen}, {Newberg}, {Nichol},
  {Nieto-Santisteban}, {Nitta}, {Oyaizu}, {Okamura}, {Ostriker}, {Padmanabhan},
  {Park}, {Peoples}, {Pier}, {Pope}, {Pourbaix}, {Quinn}, {Raddick}, {Re
  Fiorentin}, {Richards}, {Richmond}, {Rix}, {Rockosi}, {Schlegel},
  {Schneider}, {Scranton}, {Seljak}, {Sheldon}, {Shimasaku}, {Silvestri},
  {Smith}, {Smol{\v c}i{\'c}}, {Snedden}, {Stebbins}, {Stoughton}, {Strauss},
  {SubbaRao}, {Suto}, {Szalay}, {Szapudi}, {Szkody}, {Tegmark}, {Thakar},
  {Tremonti}, {Tucker}, {Uomoto}, {Vanden Berk}, {Vandenberg}, {Vidrih},
  {Vogeley}, {Voges}, {Vogt}, {Weinberg}, {West}, {White}, {Wilhite}, {Yanny},
  {Yocum}, {York}, {Zehavi}, {Zibetti}, \& {Zucker}}]{Adelman-McCarthy2007}
{Adelman-McCarthy}, J.~K. {et~al.} 2007, \apjs, 172, 634

\bibitem[{{Adelman-McCarthy} {et~al.}(2008)}]{DR6paper}
{Adelman-McCarthy}, J.~K., {et~al.} 2008, \apjs, 175, 297

\bibitem[{{Akerlof} {et~al.}(2000){Akerlof}, {Amrose}, {Balsano}, {Bloch},
  {Casperson}, {Fletcher}, {Gisler}, {Hills}, {Kehoe}, {Lee}, {Marshall},
  {McKay}, {Pawl}, {Schaefer}, {Szymanski}, \& {Wren}}]{Akerlof2000}
{Akerlof}, C. {et~al.} 2000, \aj, 119, 1901

\bibitem[{{Aldcroft} {et~al.}(2000){Aldcroft}, {Karovska},
  {Cresitello-Dittmar}, {Cameron}, \& {Markevitch}}]{aldcroft2000}
{Aldcroft}, T.~L., {Karovska}, M., {Cresitello-Dittmar}, M.~L., {Cameron},
  R.~A., \& {Markevitch}, M.~L. 2000, in Presented at the Society of
  Photo-Optical Instrumentation Engineers (SPIE) Conference, Vol. 4012, Proc.
  SPIE Vol. 4012, p. 650-657, X-Ray Optics, Instruments, and Missions III,
  Joachim E. Truemper; Bernd Aschenbach; Eds., ed. J.~E. {Truemper} \&
  B.~{Aschenbach}, 650--657

\bibitem[{{Alexander} {et~al.}(2003){Alexander}, {Bauer}, {Brandt},
  {Schneider}, {Hornschemeier}, {Vignali}, {Barger}, {Broos}, {Cowie},
  {Garmire}, {Townsley}, {Bautz}, {Chartas}, \& {Sargent}}]{alexander2003}
{Alexander}, D.~M. {et~al.} 2003, \aj, 126, 539

\bibitem[{{Amado}(2003)}]{Amado2003}
{Amado}, P.~J. 2003, \aap, 404, 631

\bibitem[{{Amado} \& {Byrne}(1997)}]{Amado1997}
{Amado}, P.~J., \& {Byrne}, P.~B. 1997, \aap, 319, 967

\bibitem[{{Anderson} {et~al.}(2007)}]{anderson06}
{Anderson}, S.~F., {et~al.} 2007, \aj, 133, 313

\bibitem[{{Apparao} {et~al.}(1992){Apparao}, {Berthiaume}, \&
  {Nousek}}]{apparao92}
{Apparao}, K.~M.~V., {Berthiaume}, G.~D., \& {Nousek}, J.~A. 1992, \apj, 397,
  534

\bibitem[{{Bade} {et~al.}(1998)}]{bade98}
{Bade}, N., {et~al.} 1998, \aaps, 127, 145

\bibitem[{{Barkhouse} {et~al.}(2006){Barkhouse}, {Green}, {Vikhlinin}, {Kim},
  {Perley}, {Cameron}, {Silverman}, {Mossman}, {Burenin}, {Jannuzi}, {Kim},
  {Smith}, {Smith}, {Tananbaum}, \& {Wilkes}}]{Barkhouse06}
{Barkhouse}, W.~A. {et~al.} 2006, \apj, 645, 955

\bibitem[{{Bochanski} {et~al.}(2007){Bochanski}, {West}, {Hawley}, \&
  {Covey}}]{Bochanski2007}
{Bochanski}, J.~J., {West}, A.~A., {Hawley}, S.~L., \& {Covey}, K.~R. 2007,
  \aj, 133, 531

\bibitem[{{Brandt} \& {Hasinger}(2005)}]{Brandt2005}
{Brandt}, W.~N., \& {Hasinger}, G. 2005, \araa, 43, 827

\bibitem[{{Cappi} {et~al.}(2001){Cappi}, {Mazzotta}, {Elvis}, {Burke},
  {Comastri}, {Fiore}, {Forman}, {Fruscione}, {Green}, {Harris}, {Hooper},
  {Jones}, {Kaastra}, {Kellogg}, {Murray}, {McNamara}, {Nicastro}, {Ponman},
  {Schlegel}, {Siemiginowska}, {Tananbaum}, {Vikhlinin}, {Virani}, \&
  {Wilkes}}]{cappi01}
{Cappi}, M. {et~al.} 2001, \apj, 548, 624

\bibitem[{{Collier Cameron} \& {Jianke}(1994)}]{Cameron1994}
{Collier Cameron}, A., \& {Jianke}, L. 1994, \mnras, 269, 1099

\bibitem[{{Covey} {et~al.}(2007){Covey}, {Ivezi{\'c}}, {Schlegel},
  {Finkbeiner}, {Padmanabhan}, {Lupton}, {Ag{\"u}eros}, {Bochanski}, {Hawley},
  {West}, {Seth}, {Kimball}, {Gogarten}, {Claire}, {Haggard}, {Kaib},
  {Schneider}, \& {Sesar}}]{Covey2007}
{Covey}, K.~R. {et~al.} 2007, \aj, 134, 2398

\bibitem[{{Cram}(1982)}]{Cram1982}
{Cram}, L.~E. 1982, \apj, 253, 768

\bibitem[{{Cutri} {et~al.}(2003)}]{Cutri2003}
{Cutri}, R.~M., {et~al.} 2003, {2MASS All Sky Catalog of point sources.} (The
  IRSA 2MASS All-Sky Point Source Catalog, NASA/IPAC Infrared Science
  Archive.~http://irsa.ipac.caltech.edu/applications/Gator/)

\bibitem[{{Della Ceca} {et~al.}(2004){Della Ceca}, {Maccacaro}, {Caccianiga},
  {Severgnini}, {Braito}, {Barcons}, {Carrera}, {Watson}, {Tedds}, {Brunner},
  {Lehmann}, {Page}, {Lamer}, \& {Schwope}}]{dellac04}
{Della Ceca}, R. {et~al.} 2004, \aap, 428, 383

\bibitem[{{Favata} {et~al.}(1992){Favata}, {Micela}, {Sciortino}, \&
  {Vaiana}}]{favata92}
{Favata}, F., {Micela}, G., {Sciortino}, S., \& {Vaiana}, G.~S. 1992, \aap,
  256, 86

\bibitem[{{Feigelson} {et~al.}(2004)}]{feigelson04}
{Feigelson}, E.~D., {et~al.} 2004, \apj, 611, 1107

\bibitem[{{Freyberg} {et~al.}(2006){Freyberg}, {Altieri}, {Bermejo}, {Esquej},
  {Lazaro}, {Read}, \& {Saxton}}]{freyberg06}
{Freyberg}, M.~J., {Altieri}, B., {Bermejo}, D., {Esquej}, M.~P., {Lazaro}, V.,
  {Read}, A.~M., \& {Saxton}, R.~D. 2006, in ESA Special Publication, Vol. 604,
  The X-ray Universe 2005, ed. A.~{Wilson}, 913--+

\bibitem[{{Fukugita} {et~al.}(1996)}]{fukugita}
{Fukugita}, M., {et~al.} 1996, \aj, 111, 1748

\bibitem[{{Gehrels}(1986)}]{gehrels86}
{Gehrels}, N. 1986, \apj, 303, 336

\bibitem[{{Gioia} {et~al.}(1984)}]{gioia84}
{Gioia}, I.~M., {et~al.} 1984, \apj, 283, 495

\bibitem[{{Gioia} {et~al.}(1990)}]{gioia90}
---. 1990, \apjs, 72, 567

\bibitem[{{Girardi} {et~al.}(2005){Girardi}, {Groenewegen}, {Hatziminaoglou}, \&
  {da Costa}}]{Girardi2005}
{Girardi}, L. {et~al.} 2005, \aap, 436, 895 

\bibitem[{{Green} {et~al.}(2004){Green}, {Silverman}, {Cameron}, {Kim},
  {Wilkes}, {Barkhouse}, {LaCluyz{\'e}}, {Morris}, {Mossman}, {Ghosh},
  {Grimes}, {Jannuzi}, {Tananbaum}, {Aldcroft}, {Baldwin}, {Chaffee}, {Dey},
  {Dosaj}, {Evans}, {Fan}, {Foltz}, {Gaetz}, {Hooper}, {Kashyap}, {Mathur},
  {McGarry}, {Romero-Colmenero}, {Smith}, {Smith}, {Smith}, {Torres},
  {Vikhlinin}, \& {Wik}}]{green04}
{Green}, P.~J. {et~al.} 2004, \apjs, 150, 43

\bibitem[{{Gunn} {et~al.}(1998)}]{gunn}
{Gunn}, J.~E., {et~al.} 1998, \aj, 116, 3040

\bibitem[{{Gunn} {et~al.}(2006)}]{gunn06}
---. 2006, \aj, 131, 2332

\bibitem[{{Hagen} {et~al.}(1995)}]{hagen95}
{Hagen}, H.-J., {et~al.} 1995, \aaps, 111, 195

\bibitem[{{Harris} \& {Johnson}(1985)}]{harris85}
{Harris}, D.~E., \& {Johnson}, H.~M. 1985, \apj, 294, 649

\bibitem[{{Harrison} {et~al.}(2003){Harrison}, {Eckart}, {Mao}, {Helfand}, \&
  {Stern}}]{harrison03}
{Harrison}, F.~A., {Eckart}, M.~E., {Mao}, P.~H., {Helfand}, D.~J., \& {Stern},
  D. 2003, \apj, 596, 944

\bibitem[{{Herbst} \& {Miller}(1989)}]{Herbst1989}
{Herbst}, W., \& {Miller}, J.~R. 1989, \aj, 97, 891

\bibitem[{{Hogg} {et~al.}(2001)}]{hogg01}
{Hogg}, D.~W., {et~al.} 2001, \aj, 122, 2129

\bibitem[{{Houdebine} {et~al.}(1996){Houdebine}, {Mathioudakis}, {Doyle}, \&
  {Foing}}]{Houdebine1996}
{Houdebine}, E.~R., {Mathioudakis}, M., {Doyle}, J.~G., \& {Foing}, B.~H. 1996,
  \aap, 305, 209

\bibitem[{{Houdebine} \& {Stempels}(1997)}]{Houdebine1997}
{Houdebine}, E.~R., \& {Stempels}, H.~C. 1997, \aap, 326, 1143

\bibitem[{{H{\"u}nsch} {et~al.}(1999){H{\"u}nsch}, {Schmitt}, {Sterzik}, \&
  {Voges}}]{Hunsch1999}
{H{\"u}nsch}, M., {Schmitt}, J.~H.~M.~M., {Sterzik}, M.~F., \& {Voges}, W.
  1999, \aaps, 135, 319

\bibitem[{{Ishisaki} {et~al.}(2001){Ishisaki}, {Ueda}, {Yamashita}, {Ohashi},
  {Lehmann}, \& {Hasinger}}]{ishisaki01}
{Ishisaki}, Y., {Ueda}, Y., {Yamashita}, A., {Ohashi}, T., {Lehmann}, I., \&
  {Hasinger}, G. 2001, \pasj, 53, 445

\bibitem[{{Ivezi{\' c}} {et~al.}(2004)}]{zeljko04}
{Ivezi{\' c}}, {\v Z}., {et~al.} 2004, Astronomische Nachrichten, 325, 583

\bibitem[{{James} {et~al.}(2000){James}, {Jardine}, {Jeffries}, {Randich},
  {Collier Cameron}, \& {Ferreira}}]{James2000}
{James}, D.~J., {Jardine}, M.~M., {Jeffries}, R.~D., {Randich}, S., {Collier
  Cameron}, A., \& {Ferreira}, M. 2000, \mnras, 318, 1217

\bibitem[{{Jardine}(2004)}]{Jardine2004}
{Jardine}, M. 2004, \aap, 414, L5

\bibitem[{{Karaali} {et~al.}(2005){Karaali}, {Bilir}, \& {Tun{\c
  c}el}}]{Karaali2005}
{Karaali}, S., {Bilir}, S., \& {Tun{\c c}el}, S. 2005, Publications of the
  Astronomical Society of Australia, 22, 24

\bibitem[{{Kim} {et~al.}(2006){Kim}, {Barkhouse}, {Romero-Colmenero}, {Green},
  {Kim}, {Mossman}, {Schlegel}, {Silverman}, {Aldcroft}, {Anderson}, {Ivezic},
  {Kashyap}, {Tananbaum}, \& {Wilkes}}]{DKim06}
{Kim}, D.-W. {et~al.} 2006, \apj, 644, 829

\bibitem[{{Kim} {et~al.}(2004{\natexlab{a}}){Kim}, {Cameron}, {Drake}, {Evans},
  {Freeman}, {Gaetz}, {Ghosh}, {Green}, {Harnden}, {Karovska}, {Kashyap},
  {Maksym}, {Ratzlaff}, {Schlegel}, {Silverman}, {Tananbaum}, {Vikhlinin},
  {Wilkes}, \& {Grimes}}]{DKim04a}
---. 2004{\natexlab{a}}, \apjs, 150, 19

\bibitem[{{Kim} {et~al.}(2004{\natexlab{b}}){Kim}, {Wilkes}, {Green},
  {Cameron}, {Drake}, {Evans}, {Freeman}, {Gaetz}, {Ghosh}, {Harnden},
  {Karovska}, {Kashyap}, {Maksym}, {Ratzlaff}, {Schlegel}, {Silverman},
  {Tananbaum}, \& {Vikhlinin}}]{DKim04b}
---. 2004{\natexlab{b}}, \apj, 600, 59

\bibitem[{{Kim} {et~al.}(2007{\natexlab{a}}){Kim}, {Kim}, {Wilkes}, {Green},
  {Kim}, {Anderson}, {Barkhouse}, {Evans}, {Ivezi{\'c}}, {Karovska}, {Kashyap},
  {Lee}, {Maksym}, {Mossman}, {Silverman}, \& {Tananbaum}}]{MKim07a}
{Kim}, M. {et~al.} 2007{\natexlab{a}}, \apjs, 169, 401

\bibitem[{{Kim} {et~al.}(2007{\natexlab{b}}){Kim}, {Wilkes}, {Kim}, {Green},
  {Barkhouse}, {Lee}, {Silverman}, \& {Tananbaum}}]{MKim07b}
{Kim}, M., {Wilkes}, B.~J., {Kim}, D.-W., {Green}, P.~J., {Barkhouse}, W.~A.,
  {Lee}, M.~G., {Silverman}, J.~D., \& {Tananbaum}, H.~D. 2007{\natexlab{b}},
  \apj, 659, 29

\bibitem[{{Kraus} \& {Hillenbrand}(2007)}]{Kraus2007}
{Kraus}, A.~L., \& {Hillenbrand}, L.~A. 2007, \aj, 134, 2340

\bibitem[{{Lavalley} {et~al.}(1992){Lavalley}, {Isobe}, \&
  {Feigelson}}]{LaValley92}
{Lavalley}, M., {Isobe}, T., \& {Feigelson}, E. 1992, in Astronomical Society
  of the Pacific Conference Series, Vol.~25, Astronomical Data Analysis
  Software and Systems I, ed. D.~M. {Worrall}, C.~{Biemesderfer}, \&
  J.~{Barnes}, 245--+

\bibitem[{{Leggett}(1992)}]{Leggett1992}
{Leggett}, S.~K. 1992, \apjs, 82, 351

\bibitem[{{Lehmann} {et~al.}(2001){Lehmann}, {Hasinger}, {Schmidt}, {Giacconi},
  {Tr{\"u}mper}, {Zamorani}, {Gunn}, {Pozzetti}, {Schneider}, {Stanke},
  {Szokoly}, {Thompson}, \& {Wilson}}]{lehmann01}
{Lehmann}, I. {et~al.} 2001, \aap, 371, 833

\bibitem[{{L{\'o}pez-Santiago} {et~al.}(2007){L{\'o}pez-Santiago}, {Micela},
  {Sciortino}, {Favata}, {Caccianiga}, {Della Ceca}, {Severgnini}, \&
  {Braito}}]{lopez07}
{L{\'o}pez-Santiago}, J., {Micela}, G., {Sciortino}, S., {Favata}, F.,
  {Caccianiga}, A., {Della Ceca}, R., {Severgnini}, P., \& {Braito}, V. 2007,
  \aap, 463, 165

\bibitem[{{Mason} {et~al.}(2000){Mason}, {Carrera}, {Hasinger}, {Andernach},
  {Aragon-Salamanca}, {Barcons}, {Bower}, {Brandt}, {Branduardi-Raymont},
  {Burgos-Mart{\'{\i}}n}, {Cabrera-Guerra}, {Carballo}, {Castander}, {Ellis},
  {Gonz{\'a}lez-Serrano}, {Mart{\'{\i}}nez-Gonz{\'a}lez},
  {Mart{\'{\i}}n-Mirones}, {McMahon}, {Mittaz}, {Nicholson}, {Page},
  {P{\'e}rez-Fournon}, {Puchnarewicz}, {Romero-Colmenero}, {Schwope}, {Vila},
  {Watson}, \& {Wonnacott}}]{mason00}
{Mason}, K.~O. {et~al.} 2000, \mnras, 311, 456

\bibitem[{{Mathioudakis} \& {Doyle}(1989)}]{Mathioudakis1989}
{Mathioudakis}, M., \& {Doyle}, J.~G. 1989, \aap, 224, 179

\bibitem[{{Monet} {et~al.}(2003){Monet}, {Levine}, {Canzian}, {Ables}, {Bird},
  {Dahn}, {Guetter}, {Harris}, {Henden}, {Leggett}, {Levison}, {Luginbuhl},
  {Martini}, {Monet}, {Munn}, {Pier}, {Rhodes}, {Riepe}, {Sell}, {Stone},
  {Vrba}, {Walker}, {Westerhout}, {Brucato}, {Reid}, {Schoening}, {Hartley},
  {Read}, \& {Tritton}}]{monet2003}
{Monet}, D.~G. {et~al.} 2003, \aj, 125, 984

\bibitem[{{Nieto--Santisteban} {et~al.}(2004)}]{sdss_images}
{Nieto--Santisteban}, M.~A., {et~al.} 2004, in ASP Conf. Ser. 314: Astronomical
  Data Analysis Software and Systems (ADASS) XIII, ed. F.~{Ochsenbein}, M.~G.
  {Allen}, \& D.~{Egret}, 666--+

\bibitem[{{Ochsenbein} {et~al.}(2000){Ochsenbein}, {Bauer}, \&
  {Marcout}}]{Ochsenbein2000}
{Ochsenbein}, F., {Bauer}, P., \& {Marcout}, J. 2000, \aaps, 143, 23

\bibitem[{{Parejko} {et~al.}(2008){Parejko}, {Constantin}, {Vogeley}, \&
  {Hoyle}}]{Parejko2008}
{Parejko}, J.~K., {Constantin}, A., {Vogeley}, M.~S., \& {Hoyle}, F. 2008, \aj,
  135, 10

\bibitem[{{Pettersen}(1987)}]{Pettersen1987}
{Pettersen}, B.~R. 1987, Vistas in Astronomy, 30, 41

\bibitem[{{Pflueger} {et~al.}(1996){Pflueger}, {Otterbein}, \&
  {Staubert}}]{pflueger96}
{Pflueger}, B., {Otterbein}, K., \& {Staubert}, R. 1996, \aap, 305, 699

\bibitem[{{Popesso} {et~al.}(2004)}]{popesso04}
{Popesso}, P., {et~al.} 2004, \aap, 423, 449

\bibitem[{{Randich} {et~al.}(1996){Randich}, {Schmitt}, \&
  {Prosser}}]{randich96}
{Randich}, S., {Schmitt}, J.~H.~M.~M., \& {Prosser}, C. 1996, \aap, 313, 815

\bibitem[{{Reid} {et~al.}(1995){Reid}, {Hawley}, \& {Mateo}}]{Reid1995}
{Reid}, N., {Hawley}, S.~L., \& {Mateo}, M. 1995, \mnras, 272, 828

\bibitem[{{Richards} {et~al.}(2007){Richards}, {Myers}, {Brunner}, {Strand},
  {Nichol}, {Gray}, {Riegel}, {Lacy}, \& {Szalay}}]{richards07}
{Richards}, G.~T. {et~al.} 2007, in American Astronomical Society Meeting
  Abstracts, Vol. 211, American Astronomical Society Meeting Abstracts,
  142.02--+

\bibitem[{{Richards} {et~al.}(2004){Richards}, {Nichol}, {Gray}, {Brunner},
  {Lupton}, {Vanden Berk}, {Chong}, {Weinstein}, {Schneider}, {Anderson},
  {Munn}, {Harris}, {Strauss}, {Fan}, {Gunn}, {Ivezi{\'c}}, {York},
  {Brinkmann}, \& {Moore}}]{richards04}
{Richards}, G.~T. {et~al.} 2004, \apjs, 155, 257

\bibitem[{{Richards} {et~al.}(2006){Richards}, {Strauss}, {Fan}, {Hall},
  {Jester}, {Schneider}, {Vanden Berk}, {Stoughton}, {Anderson}, {Brunner},
  {Gray}, {Gunn}, {Ivezi{\'c}}, {Kirkland}, {Knapp}, {Loveday}, {Meiksin},
  {Pope}, {Szalay}, {Thakar}, {Yanny}, {York}, {Barentine}, {Brewington},
  {Brinkmann}, {Fukugita}, {Harvanek}, {Kent}, {Kleinman}, {Krzesi{\'n}ski},
  {Long}, {Lupton}, {Nash}, {Neilsen}, {Nitta}, {Schlegel}, \&
  {Snedden}}]{richards06}
---. 2006, \aj, 131, 2766

\bibitem[{{Saxton} {et~al.}(2008){Saxton}, {Read}, {Esquej}, {Freyberg},
  {Altieri}, \& {Bermejo}}]{saxton2008}
{Saxton}, R.~D., {Read}, A.~M., {Esquej}, P., {Freyberg}, M.~J., {Altieri}, B.,
  \& {Bermejo}, D. 2008, \aap, 480, 611

\bibitem[{{Schmitt} {et~al.}(1985){Schmitt}, {Golub}, {Harnden}, {Maxson},
  {Rosner}, \& {Vaiana}}]{schmitt85}
{Schmitt}, J.~H.~M.~M., {Golub}, L., {Harnden}, Jr., F.~R., {Maxson}, C.~W.,
  {Rosner}, R., \& {Vaiana}, G.~S. 1985, \apj, 290, 307

\bibitem[{{Schmitt} \& {Liefke}(2004)}]{schmitt2004}
{Schmitt}, J.~H.~M.~M., \& {Liefke}, C. 2004, \aap, 417, 651

\bibitem[{{Schmitt} {et~al.}(1995)}]{schmitt95}
{Schmitt}, J.~H.~M.~M., {et~al.} 1995, \apj, 450, 392

\bibitem[{{Scranton} {et~al.}(2002){Scranton}, {Johnston}, {Dodelson},
  {Frieman}, {Connolly}, {Eisenstein}, {Gunn}, {Hui}, {Jain}, {Kent},
  {Loveday}, {Narayanan}, {Nichol}, {O'Connell}, {Scoccimarro}, {Sheth},
  {Stebbins}, {Strauss}, {Szalay}, {Szapudi}, {Tegmark}, {Vogeley}, {Zehavi},
  {Annis}, {Bahcall}, {Brinkman}, {Csabai}, {Hindsley}, {Ivezic}, {Kim},
  {Knapp}, {Lamb}, {Lee}, {Lupton}, {McKay}, {Munn}, {Peoples}, {Pier},
  {Richards}, {Rockosi}, {Schlegel}, {Schneider}, {Stoughton}, {Tucker},
  {Yanny}, \& {York}}]{Scranton2002}
{Scranton}, R. {et~al.} 2002, \apj, 579, 48

\bibitem[{{Silverman} {et~al.}(2005){Silverman}, {Green}, {Barkhouse},
  {Cameron}, {Foltz}, {Jannuzi}, {Kim}, {Kim}, {Mossman}, {Tananbaum},
  {Wilkes}, {Smith}, {Smith}, \& {Smith}}]{Silverman05}
{Silverman}, J.~D. {et~al.} 2005, \apj, 624, 630

\bibitem[{{Silvestri} {et~al.}(2006){Silvestri}, {Hawley}, {West}, {Szkody},
  {Bochanski}, {Eisenstein}, {McGehee}, {Schmidt}, {Smith}, {Wolfe}, {Harris},
  {Kleinman}, {Liebert}, {Nitta}, {Barentine}, {Brewington}, {Brinkmann},
  {Harvanek}, {Krzesi{\'n}ski}, {Long}, {Neilsen}, {Schneider}, \&
  {Snedden}}]{Silvestri2006}
{Silvestri}, N.~M. {et~al.} 2006, \aj, 131, 1674

\bibitem[{{Skrutskie} {et~al.}(1997){Skrutskie}, {Schneider}, {Stiening},
  {Strom}, {Weinberg}, {Beichman}, {Chester}, {Cutri}, {Lonsdale}, {Elias},
  {Elston}, {Capps}, {Carpenter}, {Huchra}, {Liebert}, {Monet}, {Price}, \&
  {Seitzer}}]{Skrutskie1997}
{Skrutskie}, M.~F. {et~al.} 1997, in ASSL Vol. 210: The Impact of Large Scale
  Near-IR Sky Surveys, 25--+

\bibitem[{{Skrutskie} {et~al.}(2006)}]{Skrutskie2006}
{Skrutskie}, M.~F., {et~al.} 2006, \aj, 131, 1163

\bibitem[{{Smith} {et~al.}(2002)}]{smith02}
{Smith}, J.~A., {et~al.} 2002, \aj, 123, 2121

\bibitem[{{Smol{\v c}i{\'c}} {et~al.}(2004){Smol{\v c}i{\'c}}, {Ivezi{\'c}},
  {Knapp}, {Lupton}, {Pavlovski}, {Iliji{\'c}}, {Schlegel}, {Smith}, {McGehee},
  {Silvestri}, {Hawley}, {Rockosi}, {Gunn}, {Strauss}, {Fan}, {Eisenstein}, \&
  {Harris}}]{Smolcic2004}
{Smol{\v c}i{\'c}}, V. {et~al.} 2004, \apjl, 615, L141

\bibitem[{{Stauffer} {et~al.}(1994){Stauffer}, {Caillault}, {Gagne}, {Prosser},
  \& {Hartmann}}]{Stauffer1994}
{Stauffer}, J.~R., {Caillault}, J.-P., {Gagne}, M., {Prosser}, C.~F., \&
  {Hartmann}, L.~W. 1994, \apjs, 91, 625

\bibitem[{{Stauffer} \& {Hartmann}(1986)}]{Stauffer1986}
{Stauffer}, J.~R., \& {Hartmann}, L.~W. 1986, \apjs, 61, 531

\bibitem[{{Stern} {et~al.}(2002){Stern}, {Tozzi}, {Stanford}, {Rosati},
  {Holden}, {Eisenhardt}, {Elston}, {Wu}, {Connolly}, {Spinrad}, {Dawson},
  {Dey}, \& {Chaffee}}]{stern02}
{Stern}, D. {et~al.} 2002, \aj, 123, 2223

\bibitem[{{Stocke} {et~al.}(1983)}]{stocke83}
{Stocke}, J.~T., {et~al.} 1983, \apj, 273, 458

\bibitem[{{Stocke} {et~al.}(1991)}]{stocke91}
---. 1991, \apjs, 76, 813

\bibitem[{{Stoughton} {et~al.}(2002)}]{stoughton02}
{Stoughton}, C., {et~al.} 2002, \aj, 123, 485

\bibitem[{{Sung} {et~al.}(2002){Sung}, {Bessell}, {Lee}, \& {Lee}}]{Sung2002}
{Sung}, H., {Bessell}, M.~S., {Lee}, B.-W., \& {Lee}, S.-G. 2002, \aj, 123, 290

\bibitem[{{Szkody} {et~al.}(2003)}]{paula2}
{Szkody}, P., {et~al.} 2003, \aj, 126, 1499

\bibitem[{{Szkody} {et~al.}(2005)}]{paula4}
---. 2005, \aj, 129, 2386


\bibitem[{{Vilhu}(1987)}]{Vilhu1987}
{Vilhu}, O. 1987, in Lecture Notes in Physics, Berlin Springer Verlag, Vol.
  291, Cool Stars, Stellar Systems and the Sun, ed. J.~L. {Linsky} \& R.~E.
  {Stencel}, 110--+

\bibitem[{{Vilhu} \& {Rucinski}(1983)}]{Vilhu1983}
{Vilhu}, O., \& {Rucinski}, S.~M. 1983, \aap, 127, 5

\bibitem[{{Voges} {et~al.}(1999)}]{voges99}
{Voges}, W., {et~al.} 1999, \aap, 349, 389

\bibitem[{{Voges} {et~al.}(2000)}]{fsc}
---. 2000, VizieR Online Data Catalog, 9029, 0

\bibitem[{{Walkowicz} {et~al.}(2004){Walkowicz}, {Hawley}, \&
  {West}}]{Walkowicz2004}
{Walkowicz}, L.~M., {Hawley}, S.~L., \& {West}, A.~A. 2004, \pasp, 116, 1105

\bibitem[{{Wang} {et~al.}(2004){Wang}, {Malhotra}, {Rhoads}, {Brown}, {Dey},
  {Heckman}, {Jannuzi}, {Norman}, {Tiede}, \& {Tozzi}}]{wang04}
{Wang}, J.~X. {et~al.} 2004, \aj, 127, 213

\bibitem[{{Watson} \& {XMM-Newton Survey Science Centre
  Consortium}(2006)}]{Watson2006}
{Watson}, M., \& {XMM-Newton Survey Science Centre Consortium}, t. 2006, in
  Bulletin of the American Astronomical Society, Vol.~38, Bulletin of the
  American Astronomical Society, 365--+

\bibitem[{{Weinstein} {et~al.}(2004){Weinstein}, {Richards}, {Schneider},
  {Younger}, {Strauss}, {Hall}, {Budav{\'a}ri}, {Gunn}, {York}, \&
  {Brinkmann}}]{weinstein04}
{Weinstein}, M.~A. {et~al.} 2004, \apjs, 155, 243

\bibitem[{{Weisskopf} {et~al.}(2002){Weisskopf}, {Brinkman}, {Canizares},
  {Garmire}, {Murray}, \& {Van Speybroeck}}]{weisskopf02}
{Weisskopf}, M.~C., {Brinkman}, B., {Canizares}, C., {Garmire}, G., {Murray},
  S., \& {Van Speybroeck}, L.~P. 2002, \pasp, 114, 1

\bibitem[{{West} {et~al.}(2008){West}, {Hawley}, {Bochanski}, {Covey}, {Reid},
  {Dhital}, {Hilton}, \& {Masuda}}]{West2008}
{West}, A.~A., {Hawley}, S.~L., {Bochanski}, J.~J., {Covey}, K.~R., {Reid},
  I.~N., {Dhital}, S., {Hilton}, E.~J., \& {Masuda}, M. 2008, \aj, 135, 785

\bibitem[{{Woudt} {et~al.}(2004)}]{woudt2004}
{Woudt}, P.~A., {et~al.} 2004, \mnras, 351, 1015

\bibitem[{{Wo{\'z}niak} {et~al.}(2004){Wo{\'z}niak}, {Vestrand}, {Akerlof},
  {Balsano}, {Bloch}, {Casperson}, {Fletcher}, {Gisler}, {Kehoe}, {Kinemuchi},
  {Lee}, {Marshall}, {McGowan}, {McKay}, {Rykoff}, {Smith}, {Szymanski}, \&
  {Wren}}]{wozniak04}
{Wo{\'z}niak}, P.~R. {et~al.} 2004, \aj, 127, 2436

\bibitem[{{Zickgraf} {et~al.}(2003)}]{zickgraf03}
{Zickgraf}, F.-J., {et~al.} 2003, \aap, 406, 535

\end{thebibliography}


\clearpage

\begin{landscape}
\renewcommand{\thefootnote}{\alph{footnote}}

\begin{deluxetable}{lcccccccccc}
\tablewidth{0pt}
\tabletypesize{\tiny}
\tablecaption{ChaMP Stellar Catalog (X-rays)  \label{tab:ChaMPstars-xrays}}
\tablehead{
  \colhead{Source} &
  \colhead{Chandra} &
  \colhead{fBc} &
  \colhead{netBc} &
  \colhead{fSc} &
  \colhead{netSc} &
  \colhead{HR\tablenotemark{a}} &
  \colhead{$Log (L_{Xs})$} &
  \colhead{$Log (\frac{L_{Xs}}{L_{bol}})$} &
  \colhead{$Log (L_{Xb})$} &
\colhead{$Log (\frac{L_{Xb}}{L_{bol}})$} \\
   \colhead{(CXOMP)} &
   \colhead{Obs. ID} &
   \colhead{(10$^{-13}$ ergs cm$^{-2}$ s$^{-1}$)} &
   \colhead{(counts)} &
   \colhead{(10$^{-13}$ ergs cm$^{-2}$ s$^{-1}$)} &
   \colhead{(counts)} &
   \colhead{ } &
   \colhead{(Log ergs s$^{-1}$)} &
   \colhead{ } &
   \colhead{(Log ergs s$^{-1}$)} &
\colhead{ } }
\startdata
J000155.4+004819 & 4861 &  0.08$\pm$ 0.08 &    3.5$\pm$  3.4 &  0.01$\pm$ 0.03 &    0.6$\pm$  2.3 &  1.00 & 28.60$\pm$ 1.30 & -4.44$\pm$ 1.30 & 29.60$\pm$ 1.47 & -3.45$\pm$ 1.47 \\
J001107.9+144153 & 3957 &  0.19$\pm$ 0.10 &    7.7$\pm$  4.0 &  0.10$\pm$ 0.05 &    7.9$\pm$  4.0 & -1.00 & 28.34$\pm$ 0.30 & -3.07$\pm$ 0.30 & 28.61$\pm$ 0.31 & -2.80$\pm$ 0.31 \\
J001313.2+000250 & 4829 &  0.10$\pm$ 0.05 &    8.5$\pm$  4.1 &  0.05$\pm$ 0.02 &    7.8$\pm$  4.0 & -1.00 & 29.08$\pm$ 0.31 & -4.40$\pm$ 0.31 & 29.41$\pm$ 0.29 & -4.07$\pm$ 0.29 \\
J003151.4+003233 & 2101 &  0.59$\pm$ 0.12 &   34.7$\pm$  7.0 &  0.29$\pm$ 0.06 &   33.9$\pm$  6.9 & -0.90 & 28.27$\pm$ 0.10 & -3.45$\pm$ 0.10 & 28.59$\pm$ 0.10 & -3.13$\pm$ 0.10 \\
J004238.8-091043 & 4886 &  0.97$\pm$ 0.13 &   66.4$\pm$  9.2 &  0.54$\pm$ 0.08 &   62.6$\pm$  9.0 & -0.93 & 29.73$\pm$ 0.07 & -3.11$\pm$ 0.07 & 29.98$\pm$ 0.07 & -2.86$\pm$ 0.07 \\
J010615.6+004814 & 2180 &  0.36$\pm$ 0.11 &   16.5$\pm$  5.2 &  0.18$\pm$ 0.06 &   15.9$\pm$  5.1 & -0.81 & 28.73$\pm$ 0.17 & -3.31$\pm$ 0.17 & 29.03$\pm$ 0.16 & -3.01$\pm$ 0.16 \\
J011818.5-005642 & 4963 &  0.14$\pm$ 0.03 &   36.0$\pm$  8.5 &  0.09$\pm$ 0.02 &   37.5$\pm$  7.6 & -1.00 & 29.25$\pm$ 0.10 & -3.43$\pm$ 0.10 & 29.44$\pm$ 0.12 & -3.23$\pm$ 0.12 \\
J014821.7+000446 & 4098 &  0.09$\pm$ 0.07 &    4.0$\pm$  3.2 &  0.04$\pm$ 0.03 &    3.0$\pm$  2.9 & -0.71 & 28.31$\pm$ 1.67 & -3.11$\pm$ 1.67 & 28.73$\pm$ 0.69 & -2.70$\pm$ 0.69 \\
J015939.2-084409 & 6106 &  0.01$\pm$ 0.02 &    2.2$\pm$  4.3 &  0.01$\pm$ 0.01 &    6.2$\pm$  4.0 & -1.00 & 27.06$\pm$ 0.44 & -5.57$\pm$ 0.44 & 26.85$\pm$ 1.30 & -5.79$\pm$ 1.30 \\
J015941.6-084506 & 6106 &  0.11$\pm$ 0.03 &   31.1$\pm$  7.1 &  0.07$\pm$ 0.01 &   30.5$\pm$  6.7 & -1.00 & 27.82$\pm$ 0.11 & -5.11$\pm$ 0.11 & 28.05$\pm$ 0.11 & -4.88$\pm$ 0.11 \\
J015959.7+003220 & 5777 &  0.75$\pm$ 0.08 &  115.2$\pm$ 12.6 &  0.40$\pm$ 0.04 &  103.3$\pm$ 11.5 & -0.88 & 28.91$\pm$ 0.05 & -3.57$\pm$ 0.05 & 29.18$\pm$ 0.05 & -3.29$\pm$ 0.05 \\
J020643.7+121851 & 3029 &  0.14$\pm$ 0.06 &    9.3$\pm$  4.3 &  0.06$\pm$ 0.03 &    7.8$\pm$  4.0 & -0.80 & 28.15$\pm$ 0.31 & -3.15$\pm$ 0.31 & 28.52$\pm$ 0.27 & -2.78$\pm$ 0.27 \\
J022429.5-000020 & 4987 &  0.04$\pm$ 0.01 &   43.5$\pm$  8.0 &  0.02$\pm$ 0.00 &   40.3$\pm$  7.5 & -0.91 & 27.75$\pm$ 0.09 & -3.66$\pm$ 0.09 & 28.07$\pm$ 0.09 & -3.34$\pm$ 0.09 \\
J022437.7-000711 & 4987 &  0.04$\pm$ 0.01 &   31.0$\pm$  6.9 &  0.02$\pm$ 0.00 &   31.9$\pm$  6.8 & -1.00 & 28.15$\pm$ 0.10 & -4.00$\pm$ 0.10 & 28.38$\pm$ 0.11 & -3.77$\pm$ 0.11 \\
\enddata
\footnotetext[1]{Since we do not include any scientific results based on HR, we simply characterize the typical errors here by noting that the mean error on HR is well-fit for sources with HR $>$ -0.98 by HRerr $= 0.36(\pm0.027)*HR + 0.40(\pm0.022)$, with RMS residuals of $\sigma$=0.074.  Sources with HR $<$ -0.98 have median HR errors of 0.106, with RMS residuals of $\sigma$=0.073.}
\end{deluxetable}

\begin{deluxetable}{lccccccccccc}
\tablewidth{0pt}
\tabletypesize{\tiny}
\tablecaption{ChaMP Stellar Catalog (Optical/IR)  \label{tab:ChaMPstars-oir}}
\tablehead{
  \colhead{Source} &
  \colhead{SDSS} &
  \colhead{$i$} &
  \colhead{$u-g$} &
  \colhead{$g-r$} &
  \colhead{$r-i$} &
  \colhead{$i-z$} &
  \colhead{Syn. $g$} &
  \colhead{Sat.} &
  \colhead{$J$} &
  \colhead{$J-H$} &
\colhead{$H-K_s$} \\
   \colhead{ } &
   \colhead{dist ($\arcsec$)} &
   \colhead{(mag)} &
   \colhead{(mag)} &
   \colhead{(mag)} &
   \colhead{(mag)} &
   \colhead{(mag)} &
   \colhead{(mag)} &
   \colhead{(Flag)} &
   \colhead{(mag)} &
   \colhead{(mag)} &
\colhead{(mag)} }
\startdata
J000155.4+004819 & 5.58 &   14.20$\pm$   0.01 &    1.41$\pm$   0.03 &    0.55$\pm$   0.02 &    0.20$\pm$   0.02 &    0.07$\pm$   0.03 &   15.47$\pm$   0.30 & 1 &   13.24$\pm$   0.02 &    0.38$\pm$   0.03 &    0.07$\pm$   0.04 \\
J001107.9+144153 & 0.72 &   17.16$\pm$   0.01 &    2.02$\pm$   0.23 &    1.51$\pm$   0.03 &    1.63$\pm$   0.02 &    0.89$\pm$   0.02 &   20.49$\pm$   0.30 & 0 &   14.72$\pm$   0.03 &    0.65$\pm$   0.05 &    0.26$\pm$   0.07 \\
J001313.2+000250 & 0.47 &   12.56$\pm$   0.01 &    1.17$\pm$   0.01 &    0.31$\pm$   0.01 &    0.10$\pm$   0.01 &   -0.01$\pm$   0.03 &   13.42$\pm$   0.30 & 1 &   11.73$\pm$   0.03 &    0.24$\pm$   0.04 &    0.04$\pm$   0.04 \\
J003151.4+003233 & 0.87 &   14.97$\pm$   0.03 &    2.32$\pm$   0.06 &    1.49$\pm$   0.03 &    1.46$\pm$   0.03 &    0.79$\pm$   0.03 &   17.91$\pm$   0.30 & 0 &   12.70$\pm$   0.02 &    0.56$\pm$   0.04 &    0.28$\pm$   0.04 \\
J004238.8-091043 & 0.24 &   13.36$\pm$   0.01 &    1.85$\pm$   0.07 &   -2.25$\pm$   0.07 &    3.22$\pm$   0.02 &    0.14$\pm$   0.02 &   14.82$\pm$   0.30 & 1 &   12.14$\pm$   0.02 &    0.45$\pm$   0.04 &    0.13$\pm$   0.04 \\
J010615.6+004814 & 0.23 &   15.75$\pm$   0.03 &    2.48$\pm$   0.08 &    1.40$\pm$   0.04 &    1.24$\pm$   0.04 &    0.67$\pm$   0.03 &   18.26$\pm$   0.30 & 0 &   13.61$\pm$   0.03 &    0.60$\pm$   0.04 &    0.23$\pm$   0.04 \\
J011818.5-005642 & 1.61 &   15.28$\pm$   0.02 &    2.35$\pm$   0.04 &    1.15$\pm$   0.04 &    0.52$\pm$   0.03 &    0.31$\pm$   0.03 &   17.25$\pm$   0.30 & 0 &   13.72$\pm$   0.02 &    0.61$\pm$   0.04 &    0.16$\pm$   0.04 \\
J014821.7+000446 & 1.10 &   18.10$\pm$   0.01 &    2.40$\pm$   0.80 &    1.62$\pm$   0.05 &    1.58$\pm$   0.03 &    0.82$\pm$   0.03 &   20.70$\pm$   0.30 & 0 &   15.77$\pm$   0.07 &    0.76$\pm$   0.10 &    0.18$\pm$   0.13 \\
J015939.2-084409 & 1.40 &   15.18$\pm$   0.02 &    2.64$\pm$   0.03 &   -0.96$\pm$   0.04 &   -0.54$\pm$   0.04 &    3.62$\pm$   0.02 &   13.76$\pm$   0.30 & 1 &   10.41$\pm$   0.02 &    0.62$\pm$   0.04 &    0.16$\pm$   0.03 \\
J015941.6-084506 & 0.33 &   10.18$\pm$   0.01 &    1.87$\pm$   0.01 &    0.82$\pm$   0.01 &    0.30$\pm$   0.01 &   -1.97$\pm$   0.02 &   11.82$\pm$   0.30 & 1 &    9.34$\pm$   0.02 &    0.41$\pm$   0.06 &    0.09$\pm$   0.06 \\
J015959.7+003220 & 1.32 &   13.86$\pm$   0.01 &    2.24$\pm$   0.03 &    1.43$\pm$   0.02 &    1.36$\pm$   0.01 &    0.73$\pm$   0.02 &   15.42$\pm$   0.30 & 1 &   11.72$\pm$   0.02 &    0.64$\pm$   0.03 &    0.25$\pm$   0.03 \\
J020643.7+121851 & 0.52 &   17.59$\pm$   0.02 &    2.45$\pm$   0.50 &    1.63$\pm$   0.04 &    1.61$\pm$   0.03 &    0.88$\pm$   0.02 &   20.84$\pm$   0.30 & 0 &   15.20$\pm$   0.04 &    0.65$\pm$   0.06 &    0.39$\pm$   0.08 \\
J022429.5-000020 & 0.43 &   17.50$\pm$   0.02 &    2.62$\pm$   0.56 &    1.46$\pm$   0.05 &    1.67$\pm$   0.03 &    0.87$\pm$   0.03 &   20.45$\pm$   0.30 & 0 &   15.15$\pm$   0.04 &    0.68$\pm$   0.06 &    0.31$\pm$   0.08 \\
J022437.7-000711 & 0.85 &   16.14$\pm$   0.02 &    2.71$\pm$   0.12 &    1.41$\pm$   0.03 &    1.06$\pm$   0.03 &    0.58$\pm$   0.03 &   18.94$\pm$   0.30 & 0 &   14.10$\pm$   0.03 &    0.57$\pm$   0.04 &    0.27$\pm$   0.05 \\
\enddata
\end{deluxetable}

\clearpage
\end{landscape}

\begin{deluxetable}{lccc}
\tablewidth{0pt}
\tabletypesize{\scriptsize}
\tablecaption{ChaMP Stars With Spectra. \label{tab:ChaMPstars-specs}}
\tablehead{
\colhead{} &
\colhead{} &
\colhead{H$\alpha$ EqW} &
\colhead{log } \\
\colhead{CXOMP} &
\colhead{Type} &
\colhead{(\AA)} &
\colhead{L$_{\rm H_\alpha}$/L$_{\rm bol}$} 
}
\startdata
J001107.9$+$144153 & M5 & $+$6.09 & $-$3.78 \\
J001313.2$+$000250 & F7 & $-$3.67 & \nodata \\
J003151.4$+$003233 & M4 & $+$5.01 & $-$3.50 \\
J010615.6$+$004814 & M3 & $+$2.71 & $-$3.70 \\
J011818.5$-$005642 & K5 & $+$0.71 & \nodata \\
J015941.6$-$084506\tablenotemark{a} & K2 & $-$1.76 & \nodata \\
J020643.7$+$121851 & M5 & $+$6.07 & $-$3.78 \\
J022429.5$-$000020 & M4 & $+$3.43 & $-$3.66 \\
J023206.6$-$073032 & M4 & $+$2.75 & $-$3.76 \\
J025951.7$+$004619\tablenotemark{b} & M5 & $+$14.47 & $-$3.40 \\
J030014.0$+$004729 & K1 & $-$1.52 & \nodata \\
J072501.4$+$371351\tablenotemark{a} & K3 & $-$1.15 & \nodata \\
J074108.8$+$311346\tablenotemark{a} & M6 & $+$1.24 & $-$4.66 \\
J074112.3$+$311446 & M4 & $+$8.60 & $-$3.26 \\
J074118.8$+$311434 & M3 & $+$3.53 & $-$3.58 \\
J074433.8$+$393027 & F2 & $-$3.55 & \nodata \\
J074437.0$+$392503 & M5 & $+$5.71 & $-$3.81 \\
J074444.6$+$392931 & F5 & $-$3.69 & \nodata \\
J074705.1$+$274006 & G1 & $-$2.53 & \nodata \\
J075549.9$+$405728 & M2 & $+$0.09 & $-$5.08 \\
J075937.3$+$300846 & G7 & $-$0.88 & \nodata \\
J080046.6$+$360416 & M2 & $-$0.05 & \nodata \\
J080048.1$+$360722 & F1 & $-$3.54 & \nodata \\
J080101.0$+$360549 & G2 & $-$2.57 & \nodata \\
J080157.1$+$441438 & K2 & $-$0.74 & \nodata \\
J082702.2$+$291531 & G4 & $-$0.94 & \nodata \\
J082718.9$+$291841 & M0 & $-$0.45 & \nodata \\
J082726.2$+$291601 & G2 & $-$1.50 & \nodata \\
J082729.8$+$291905 & K2 & $-$1.11 & \nodata \\
J082815.2$+$291132 & G5 & $-$2.25 & \nodata \\
J084039.0$+$130916 & M0 & $-$0.30 & \nodata \\
J084044.7$+$130713\tablenotemark{a} & G1 & $-$2.78 & \nodata \\
J084055.8$+$130800 & M4 & $+$2.97 & $-$3.72 \\
J084913.8$+$444758 & M0 & $-$0.42 & \nodata \\
J085325.7$+$232919\tablenotemark{a} & M4 & $+$6.47 & $-$3.39 \\
J091047.6$+$541505 & M2 & $+$2.62 & $-$3.60 \\
J091104.1$+$542208 & M2 & $-$0.52 & \nodata \\
J093411.0$+$551143 & F4 & $-$4.09 & \nodata \\
J111504.7$+$403706 & M3 & $+$1.04 & $-$4.12 \\
J111802.3$+$074325 & G0 & $-$2.88 & \nodata \\
J112045.7$+$232536 & M2 & $+$3.67 & $-$3.45 \\
J112116.2$+$232622 & M3 & $+$0.17 & $-$4.90 \\
J114007.3$+$660659 & M2 & $+$2.43 & \nodata \\
J114101.7$+$661246 & M3 & $+$0.80 & $-$4.23 \\
J114129.7$+$660250 & K3 & $-$1.29 & \nodata \\
J114149.5$+$661123 & K7 & $-$0.31 & \nodata \\
J120439.4$-$001650 & G8 & $-$1.98 & \nodata \\
J122155.2$+$490743 & M4 & $+$4.79 & $-$3.52 \\
J122738.8$+$442132 & M4 & $+$3.28 & $-$3.68 \\
J122837.1$+$015720\tablenotemark{b} & M1 & $+$1.19 & \nodata \\
J125152.2$+$000528 & M3 & $+$5.23 & $-$3.41 \\
J131231.0$+$423106 & M3 & $+$2.47 & $-$3.74 \\
J134433.5$-$000536 & M4 & $+$7.23 & $-$3.34 \\
J134434.8$+$554956 & M3 & $+$3.95 & $-$3.53 \\
J134449.1$+$555812 & F8 & $-$1.81 & \nodata \\
J134521.5$-$000118 & M3 & $+$2.43 & $-$3.75 \\
J140654.3$+$340949 & M4 & $+$3.81 & $-$3.62 \\
J141120.7$+$521411 & K7 & $-$0.94 & \nodata \\
J141715.2$+$445420 & G7 & $-$1.86 & \nodata \\
J144553.5$+$012552 & M0 & $+$0.55 & $-$4.20 \\
J150639.4$+$521856 & M6 & $+$3.43 & $-$4.22 \\
J151031.7$+$074248 & K2 & $-$0.07 & \nodata \\
J151423.8$+$363511 & M3 & $+$1.35 & $-$4.00 \\
J153245.2$-$004012 & G8 & $-$1.97 & \nodata \\
J153519.7$+$233152 & M2 & $-$0.32 & \nodata \\
J154905.1$+$213319 & K3 & $-$1.23 & \nodata \\
J154947.2$+$212857 & G3 & $-$2.57 & \nodata \\
J161958.8$+$292321 & M2 & $+$3.56 & $-$3.46 \\
J162306.8$+$311236 & M4 & $+$3.85 & $-$3.61 \\
J162415.4$+$263728 & K2 & $-$0.90 & \nodata \\
J214218.8$+$122524 & G2 & $-$1.72 & \nodata \\
J214229.3$+$123317 & M2 & $+$3.63 & $-$3.45 \\
J214229.3$+$123322 & M4 & $+$13.54 & $-$3.06 \\
J214235.6$+$122701 & K0 & $-$1.83 & \nodata \\
J221513.1$-$004828 & M2 & $+$1.39 & $-$3.87 \\
J221513.2$-$004927 & G4 & $-$2.77 & \nodata \\
J221516.8$-$005129 & M4 & $+$5.51 & $-$3.46 \\
J221716.9$+$002208 & K5 & $-$1.04 & \nodata \\
J221719.1$+$001428 & K4 & $+$0.41 & \nodata \\
J224339.9$-$093348 & K3 & $-$0.65 & \nodata \\
J231818.7$+$003842 & K7 & $-$0.96 & \nodata \\
J231820.3$+$003129\tablenotemark{b} & F2 & $-$3.78 & \nodata \\
\enddata
\tablenotetext{1}{Counterpart is cataloged in SIMBAD but lacks a stellar type.}
\tablenotetext{2}{Counterpart is cataloged in SIMBAD and has a stellar type.}
\end{deluxetable}

\begin{deluxetable}{llcccl}
\tablewidth{0pt}
\tabletypesize{\tiny}
\tablecaption{ChaMP Sources With SIMBAD Counterparts. \label{tab:simbad-stars}}
\tablehead{
  \colhead{} &
  \colhead{SIMBAD} &
  \colhead{Sep.} &
  \colhead{$B$} &
  \colhead{$V$} &
  \colhead{} \\
  \colhead{CXOMP} &
  \colhead{Counterpart} &
  \colhead{(\arcsec)} &
  \colhead{(mag)} &
  \colhead{(mag)} &
  \colhead{Comments}
}
\startdata
J015941.6$-$084506 & BD$-$09 375  & $0.4$ & $11.90$ & $10.80$ & \nodata  \\
J015959.7$+$003220 & [BHR2005] 829$-$29  & $1.8$ & \nodata & \nodata & M3.5V \\
J023132.5$-$072724 & TYC 4704$-$81$-$1  & $1.1$ & $11.70$ & $11.20$ & \nodata \\
J025951.7$+$004619 & [BHR2005] 832$-$7  & $0.0$ & \nodata & \nodata & M5.5V \\
J072501.4$+$371351 & TYC 2464$-$396$-$1  & $0.9$ & $10.55$ & $9.52$ & \nodata \\
J072545.4$+$365905 & BD$+$37 1715  & $0.6$ & $10.72$ & $10.28$ & F5  \\
J074108.8$+$311346 & 2MASS J07410881$+$3113463  & $0.0$ & $19.20$ & \nodata & \nodata \\
J080500.8$+$103001 & HD 66686  & $1.9$ & $8.17$ & $7.29$ & G5III \\
J080853.8$+$201641 & BD$+$20 2009  & $2.3$ & $10.86$ & $10.60$ & F5 \\
J080920.5$+$202322 & BD$+$20 2011  & $1.2$ & $10.82$ & $10.14$ & F0 \\
J081539.2$+$364742 & TYC 2482$-$1192$-$1  & $1.7$ & $11.70$ & $11.10$ & \nodata \\
J084044.7$+$130713 & TYC 805$-$471$-$1  & $1.2$ & $11.80$ & $11.20$ & \nodata \\
J084944.7$+$445840\tablenotemark{a,b} & HD 75117 & $0.8$ & $8.45$ & $7.98$ & G0 \\
J085005.3$+$445819\tablenotemark{b} & HD 75172 & $4.2$ & $9.24$ & $8.98$ & F5 \\
J085318.2$+$281106 & TYC 1949$-$1327$-$1  & $1.5$ & $11.20$ & $10.39$ & \nodata \\
J085325.7$+$232919 & 2MASS J08532577$+$2329194 & $1.1$ & \nodata & \nodata & \nodata \\ 
J085711.0$+$085651 & TYC 811$-$1921$-$1  & $2.6$ & $11.10$ & $10.70$ & \nodata \\
J090941.7$+$541939\tablenotemark{b} & TYC 3805$-$167$-$1  & $2.5$ & $10.95$ & $10.50$ & \nodata \\
J091432.7$+$561238 & HD 237796  & $0.6$ & $10.65$ & $9.61$ & K5 \\
J091444.7$+$562104 & TYC 3809$-$904$-$1  & $1.8$ & $11.40$ & $10.90$ & \nodata \\
J093342.5$+$340154 & TYC 2497$-$1154$-$1  & $0.8$ & $11.70$ & $11.30$ & \nodata \\
J093905.4$+$005146 & TYC 235$-$1240$-$1  & $2.1$ & $11.50$ & $11.00$ & \nodata \\
J095427.2$+$410515 & BD$+$41 2023  & $1.7$ & $11.10$ & $10.36$ & K0 \\
J095721.0$+$465821 & TYC 3433$-$1205$-$1  & $2.0$ & $11.60$ & $10.90$ & \nodata \\
J103857.9$+$400335 & TYC 3005$-$806$-$1  & $1.3$ & $10.50$ & $10.22$ & \nodata \\
J104320.6$+$005954 & TYC 254$-$1148$-$1  & $3.3$ & $11.90$ & $11.10$ & \nodata \\
J105202.9$+$160544 & CCDM J10520$+$1606AB  & $0.6$ & $8.68$ & $8.31$ & F5 ** \\ 
J105211.8$+$161002 & BD$+$16 2181  & $0.9$ & $9.93$ & $9.66$ & F5 \\	
J111446.9$+$532038 & TYC 3824$-$287$-$1  & $1.2$ & $11.19$ & $10.71$ & \nodata \\
J111548.9$+$532234 & 2MASS J11154905$+$5322345  & $1.4$ & $12.80$ & \nodata & \nodata \\
J111607.3$+$013512 & CCDM J11161$+$0135AB  & $0.8$ & $9.48$ & $9.01$ & F8 ** \\
J111607.3$+$013509 & CCDM J11161$+$0135AB  & $3.2$ & $9.48$ & $9.01$ & F8 ** \\ 
J114144.3$+$654114 & HD 101557  & $1.5$ & $9.65$ & $8.97$ & G5 \\
J115931.2$+$553109 & GPM 179.879716$+$55.519459  & $1.2$ & $12.00$ & $11.50$ & \nodata \\
J120041.3$+$290512 & BD$+$29 2244  & $4.1$ & $11.08$ & $10.57$ & F6 \\
J120154.0$+$575636 & HD 104482  & $0.9$ & $9.48$ & $9.02$ & F5 \\
J120205.1$+$575539 & HD 238063  & $0.4$ & $10.37$ & $9.97$ & F5 \\
J124533.5$+$005914 & HD 110935  & $1.5$ & $9.98$ & $9.08$ & G5 \\
J125152.0$+$000505 & HD 111816  & $1.0$ & $8.31$ & $7.83$ & F8 \\
J125605.1$+$260117 & BD$+$26 2407  & $2.3$ & $10.09$ & $9.73$ & F2 \\
J125615.7$+$564817 & GSC 03845$-$00748  & $1.2$ & $12.50$ & \nodata & \nodata \\
J130115.9$+$002958 & NLTT 32614  & $2.6$ & $10.45$ & $9.59$ & K0 PM* \\
J130549.6$+$035341 & LSPM J1305$+$0353  & $0.4$ & $18.80$ & \nodata & PM* \\
J130908.5$+$212721 & CCDM J13091$+$2127AB  & $0.9$ & $8.93$ & $8.34$ & G5 ** \\
J130955.3$+$573403 & HD 114505  & $1.1$ & $9.80$ & $9.21$ & G \\
J131043.9$-$031731  & HD 114465  & $0.7$ & $9.12$ & $8.71$ & F8 \\
J131057.7$+$011553 & StKM 1$-$1052  & $3.2$ & \nodata & $11.40$ & K5 \\
J133143.4$+$111132 & BD$+$11 2580  & $0.7$ & $11.00$ & $9.96$ & K0 \\
J141345.7$+$000710 & 2MASS J14134569$+$0007068  & $2.8$ & $18.96$ & \nodata & \nodata \\
J141349.2$+$000806 & 2MASS J14134944$+$0008055  & $3.5$ & $15.97$ & \nodata & \nodata \\
J141808.1$+$264743 & HD 125320  & $0.8$ & $8.93$ & $8.19$ & G5IV \\	
J143800.1$+$033528 & HD 128645  & $0.5$ & $9.72$ & $9.40$ & F5 \\
J144232.8$+$011710 & BD$+$01 2965  & $0.7$ & $10.41$ & $9.10$ & K0III \\
J144300.8$+$012423 & 2MASS J14430071$+$0124239  & $1.8$ & $15.30$ & \nodata & \nodata \\
J144430.1$-$012826  & 2MASS J14443011$-$0128261  & $0.3$ & $15.30$ & $15.29$ & RR* \\
J144848.7$+$474041 & GSC 03477$-$01108  & $1.1$ & $14.00$ & \nodata & \nodata \\
J152209.8$+$524435 & HD 137146  & $1.9$ & $9.70$ & $9.26$ & F8 	\\
J153203.6$+$240505 & BD$+$24 2880  & $0.8$ & $9.86$ & $9.42$ & F8 \\
J153658.6$+$120011 & GPM 234.244393$+$12.002844  & $0.7$ & $12.40$ & $11.30$ & \nodata \\
J160234.6$+$423021 & HD 144129  & $0.6$ & $9.85$ & $9.46$ & F5 \\
J162157.2$+$381734\tablenotemark{c} & TYC 3062$-$1317$-$1  & $0.2$ & $11.80$ & $10.50$ & Ce*\tablenotemark{d} \\
J162357.2$+$311253 & TYC 2580$-$796$-$1  & $0.7$ & $11.70$ & $11.50$ & \nodata \\ 
J170137.9$+$400548 & TYC 3076$-$801$-$1  & $3.0$ & $10.93$ & $10.11$ & \nodata \\
J171952.9$+$263003 & NLTT 44615  & $1.0$ & $13.80$ & $13.60$ & PM* \\
J231820.3$+$003129 & HD 219752  & $1.1$ & $9.54$ & $9.11$ & F5 \\
J231956.4$+$003418 & TYC 577$-$673$-$1  & $1.0$ & $11.90$ & $11.20$ & \nodata \\
\enddata
\tablenotetext{1}{Source cataloged in \citet{stern02}.}
\tablenotetext{2}{Source cataloged in \citet{harrison03}.}
\tablenotetext{3}{Source cataloged in \citet{fsc}.}
\tablenotetext{4}{While this star is classified as a Cepheid by \citet{Akerlof2000}, inspection of the Northern Sky Variability Survey \citep{wozniak04} lightcurve for this object suggests that this is not a classical Cepheid.}
\tablecomments{``**'' indicates a double or multiple star, ``PM*'' indicates a high proper motion stars, ``RR*'' a RR Lyr-type variable, and ``Ce*'' a Cepheid-type variable.}    
\end{deluxetable}


\begin{deluxetable}{llcl}
\tablewidth{0pt}
\tabletypesize{\tiny}
\tablecaption{ChaMP Sources Included In Other X-ray Catalogs. \label{tab:other-xray}}
\tablehead{
  \colhead{CXOMP} &
  \colhead{Other Name} &
  \colhead{Sep.\ (\arcsec)} &
  \colhead{Catalog}
}
\startdata
J084854.0$+$450230 & [STS2002] 43                & $1.1$ & \citet{stern02} \\
 		   & CXOSEXSI J084854.0$+$450231 & $1.4$ & \citet{harrison03} \\
J084913.8$+$444758 & [STS2002] 88                & $1.8$ & \citet{stern02} \\
J084921.3$+$444949 & CXOSEXSI J084921.2$+$444948 & $0.5$ & \citet{harrison03} \\
		   & [STS2002] 106               & $0.9$ & \citet{stern02} \\
J091045.7$+$542019 & CXOSEXSI J091045.7$+$542019 & $2.9$ & \citet{harrison03} \\
J091047.6$+$541505 & CXOSEXSI J091047.6$+$541505 & $0.8$ & \citet{harrison03} \\
J091104.1$+$542208 & CXOSEXSI J091104.2$+$542206 & $1.9$ & \citet{harrison03} \\
J115903.8$+$291747 & CXOSEXSI J115903.7$+$291746 & $1.3$ & \citet{harrison03} \\
J125152.2$+$000528 & [ZEH2003] RX J1251.8$+$0005 1 & $3.2$ & \citet{zickgraf03} \\
J141120.7$+$521411 & CXOSEXSI J141120.7$+$521411 & $0.2$ & \citet{harrison03} \\
		   & [CME2001] 3C 295 12  	 & $1.0$ & \citet{cappi01} \\
		   & [ZEH2003] RX J1411.3$+$5212 1 & $1.4$ & $B = 14.1$; \citet{zickgraf03} \\ 
J142527.4$+$352656 & CXOLALA1 J142527.5$+$352656 & $1.6$ & \citet{wang04} \\
J142547.1$+$353954 & CXOLALA1 J142547.1$+$353954 & $0.9$ & \citet{wang04} \\
J162415.4$+$263728 & CXOSEXSI J162415.4$+$263729 & $1.2$ & \citet{harrison03} \\
J214229.3$+$123322 & 1RXS J214229.5$+$123323     & $3.3$ & \citet{fsc} \\
\enddata
\end{deluxetable}

\begin{deluxetable}{llcccl}
\tablewidth{0pt}
\tabletypesize{\tiny}
\tablecaption{Previously Known Stellar X-ray Sources With ChaMP Detections. \label{tab:known-sources}}
\tablehead{
  \colhead{} &
  \colhead{SIMBAD} &
  \colhead{Sep.} &
  \colhead{$B$} &
  \colhead{$V$} &
  \colhead{} \\
  \colhead{CXOMP} &
  \colhead{Counterpart} &
  \colhead{(\arcsec)} &
  \colhead{(mag)} &
  \colhead{(mag)} &
  \colhead{Comments}
}
\startdata
J080813.5$+$210608 & LHS 5134  & $0.5$ & $12.65$ & $11.20$ & M2.5 PM*; \citet{Hunsch1999} \\
J100734.8$+$130144 & NLTT 23457  & $3.6$ & $8.98$ & $8.40$ & F8 PM*; \citet{stocke91}  \\ 
J105336.4$+$573800 & TYC 3829$-$162$-$1 &  $0.6$ & $11.00$ & $10.43$ & \citet{ishisaki01}\tablenotemark{a} \\
J105410.3$+$573038 & RDS 20C 	& $0.8$ & \nodata & \nodata & M5V; \citet{ishisaki01} \\
J122156.1$+$271834 & HD 107611  & $0.9$ & $8.95$ & $8.50$ & F6V *iC; \citet{randich96} \\
J122837.1$+$015720\tablenotemark{b} & GSC 00282$-$00187  & $3.6$ & \nodata & \nodata & M2; \citet{pflueger96} \\
J125533.7$+$255331 & PN G339.9$+$88.4  & $0.8$ & $9.65$ & $8.86$ & G5III PN; \citet{apparao92} \\
J134513.3$+$555244 & NLTT 35142  & $1.8$ & $6.97$ & $6.50$ & F7IV$-$V PM*; \citet{schmitt85} \\
J135608.7$+$183039 & GSC 01470$-$00791  & $1.4$ & \nodata & $15.00$ & K5; \citet{mason00} \\
J171954.1$+$263003 & V* V647 Her  & $1.4$ & $12.98$ & $11.42$ & M4 Fl*; \citet{harris85} \\
\enddata
\tablenotetext{1}{Both TYC 3829$-$162$-$1 and its probable X-ray counterpart, RX J105336.4$+$573802 (detected by \citet{lehmann01}), are listed in \citet{ishisaki01}. This is presumably an accounting error.}
\tablenotetext{2}{Also included in the \citet{lopez07} sample.}
\tablecomments{``PM*'' indicates a high proper motion stars, ``*iC'' a star in a cluster, ``PN'' a planetary nebula, ``Ce*'' a Cepheid-type variable, and ``Fl*'' a flare star.}    
\end{deluxetable}



\end{document}